\documentclass[a4paper,fleqn,usenatbib]{mnras}

\usepackage{newtxtext,newtxmath}

\usepackage[T1]{fontenc}
\usepackage{ae,aecompl}

\usepackage{graphicx}	
\usepackage{amsmath}	
\usepackage{wasysym}

\def\mrpd{\hbox{mrad\,d$^{-1}$}}

\def\chisqr{\hbox{$\chi^2_{\rm r}$}}
\def\msun{\hbox{${\rm M}_{\odot}$}}
\def\mjup{\hbox{${\rm M}_{\jupiter}$}}

\def\mspy{\hbox{${\rm M}_{\odot}$\,yr$^{-1}$}}
\def\rsun{\hbox{${\rm R}_{\odot}$}}
\def\lsun{\hbox{${\rm L}_{\odot}$}}
\def\rcor{\hbox{$r_{\rm cor}$}}
\def\rmag{\hbox{$r_{\rm mag}$}}
\def\mstar{\hbox{$M_{\star}$}}
\def\rstar{\hbox{$R_{\star}$}}
\def\lstar{\hbox{$L_{\star}$}}
\def\teff{\hbox{$T_{\rm eff}$}}
\def\logg{\hbox{$\log g$}}

\def\vD{\hbox{$v_{\rm D}$}}

\def\kms{\hbox{km\,s$^{-1}$}}

\def\vsini{\hbox{$v \sin i$}}

\def\mic{\hbox{$\mu$m}}

\def\emr{}

\def\Bl{\hbox{$B_{\rm \ell}$}}
\def\Bd{\hbox{$B_{\rm d}$}}

\def\degr{\hbox{$^\circ$}}

\def\Mdot{\hbox{$\dot{M}$}}

\def\Prot{\hbox{$P_{\rm rot}$}}

\newcommand{\caii}{\hbox{Ca$\;${\sc ii}}}

\newcommand{\hei}{\hbox{He$\;${\sc i}}}
\newcommand{\hal}{\hbox{H${\alpha}$}}
\newcommand{\hbe}{\hbox{H${\beta}$}}
\newcommand{\pab}{\hbox{Pa${\beta}$}}
\newcommand{\brg}{\hbox{Br${\gamma}$}}

\title[Magnetic field, accretion and planets of CI~Tau]{The classical T~Tauri star CI~Tau observed with SPIRou: magnetospheric accretion and planetary formation} 
\author[J.-F.~Donati et al.]{J.-F.~Donati$^{1}$\thanks{E-mail: jean-francois.donati@irap.omp.eu},
           B.~Finociety$^{1}$, P.I.~Cristofari$^{1,2}$, S.H.P.~Alencar$^{3}$, C.~Moutou$^1$,  
\newauthor X.~Delfosse$^{4}$, P Fouqu\'e$^{1}$, L.~Arnold$^{5}$, C.~Baruteau$^{1}$, \'A.~K\'osp\'al$^{6}$,  
\newauthor F.~M\'enard$^{4}$, A.~Carmona$^{4}$, K.~Grankin$^{7}$, M.~Takami$^{8}$, E.~Artigau$^{9}$, 
\newauthor R.~Doyon$^{9}$, G.~H\'ebrard$^{10}$ and the SLS collaboration  
\vspace{1mm}
\\ 
$^1$ Univ.\ de Toulouse, CNRS, IRAP, 14 avenue Belin, 31400 Toulouse, France \\ 
$^2$ Center for Astrophysics, Harvard \& Smithsonian, 60 Garden street, Cambridge, MA 02138, United States \\
$^3$ Departamento de F\'{\i}sica -- ICEx -- UFMG, Av. Ant\^onio Carlos, 6627, 30270-901 Belo Horizonte, MG, Brazil\\  
$^4$ Univ.\ Grenoble Alpes, CNRS, IPAG, 38000 Grenoble, France \\ 
$^5$ Canada-France-Hawaii Telescope, 65-1238 Mamalahoa Hwy., Kamuela, HI 96743, USA \\ 
$^6$ Konkoly Observatory, Research Centre for Astronomy and Earth Sciences, Konkoly-Thege Mikl\'os \'ut 15-17, 1121 Budapest, Hungary \\
$^7$ Crimean Astrophysical Observatory, Nauchny, Crimea 298409 \\ 
$^8$ Institute of Astronomy and Astrophysics, Academia Sinica, Roosevelt Rd, Taipei 10617, Taiwan \\ 
$^9$ Universit\'e de Montr\'eal, D\'epartement de Physique, IREX, Montr\'eal, QC H3C 3J7, Canada \\ 
$^{10}$ Institut d'Astrophysique de Paris, CNRS, Sorbonne Univ., 98 bis bd Arago, 75014 Paris, France  
}

\date{Submitted 2023 September 30 -- Accepted 2024 February 29} 

\pubyear{2023}

\begin{document}

\label{firstpage}
\pagerange{\pageref{firstpage}--\pageref{lastpage}}
\maketitle

\begin{abstract}
{\emr 
We report new observations of the classical T~Tauri star CI~Tau with the SPIRou near-infrared spectropolarimeter and velocimeter at the Canada-France-Hawaii Telescope (CFHT) in late 2019, 2020 and 2022, 
complemented with observations obtained with the ESPaDOnS optical spectropolarimeter at CFHT in late 2020.  From our SPIRou and ESPaDOnS spectra, to which we applied Least-Squares Deconvolution, 
we infer longitudinal fields clearly modulated with the 9-d rotation period of CI~Tau.  
Using Zeeman-Doppler imaging, we reconstruct the large-scale magnetic topology, first from SPIRou data only in all three seasons, then from our 2020 SPIRou and ESPaDOnS data simultaneously.  
We find that CI~Tau hosts a mainly axisymmetric poloidal field, with a 1~kG dipole slightly tilted to the rotation axis and dark spots close to the pole that coincide with the footpoints 
of accretion funnels linking the star to the inner disc.  Our results also suggest that CI~Tau accretes mass from the disc in a stable fashion.  
We further find that radial velocities (RV) derived from atomic and CO lines in SPIRou spectra are both rotationally modulated, but with a much lower amplitude than that expected from the putative  
candidate planet CI~Tau~b.  We confirm the presence of a RV signal at a period of 23.86~d reported in a separate analysis, but detect it clearly in CO lines only and not in atomic lines, suggesting that 
it likely traces a non-axisymmetric structure in the inner disc of CI~Tau rather than a massive close-in planet. } 
\end{abstract}




\section{Introduction}
\label{sec:int}

In the last three decades, thousands of extra-solar planets and exoplanet systems have been detected and characterized by various methods, in particular velocimetry, measuring planet masses through the 
reflex motion planets induce on their host stars, and photometry, measuring planet radii from the flux reduction planets generate as they transit in front of their host stars.  However, very few observations 
exist about nascent exoplanets, younger than, e.g., 20~Myr, with only a couple of multi-planet systems known to date \citep[e.g.,][]{David19,Plavchan20,Donati23,Finociety23b}.  Yet, documenting 
such objects is key to provide constraints on theoretical models of how planets build up and migrate in the protoplanetary accretion disc surrounding forming stars, and from which both stars and planets feed.  

Pre-main-sequence (PMS) low-mass stars, also called T~Tauri stars (TTSs), are particularly interesting in this respect, both those that are still surrounded by an accretion disc (the classical TTSs / cTTSs) 
and the ones that exhausted their discs (the weak-line TTSs / wTTSs).  Whereas cTTSs exhibit extended discs with a wide variety of spatial structures, sometimes even featuring distant planets detected through 
imaging techniques \citep{Keppler18}, wTTSs are known to host close-in massive planets, some of them even transiting in front of their host stars \citep{David16,David19}, giving the opportunity to derive 
estimates of the planet bulk densities and thereby hints on the planet internal structures.  

Magnetic fields are known to play a key role throughout the whole process of star and planet formation \citep[e.g.,][]{Hennebelle20}, including in the cTTS phase where the dynamo-generated large-scale field 
of the host star is strong enough to evacuate the central regions of the accretion disc and carve a large magnetospheric gap extending up to 10 stellar radii.  The magnetic field of the central star connects 
to that at the inner edge of the accretion disc, controlling accretion between the disc and the star and potentially slowing down the rotation of the host star \citep[e.g.,][]{Romanova04,Zanni13,Blinova16,Pantolmos20}.  
This star-disc interaction may also impact the fate of newborn planets as they migrate inwards within the accretion disc and pile up near its inner edge \citep[e.g.,][]{Mulders15}, allowing them to potentially 
escape from being swallowed by the host star \citep{Lin96,Romanova06}.  

One such cTTS of particular interest is CI~Tau, aged 2~Myr and known to host a massive accretion disc extending to several hundreds of au from the central star \citep[e.g.,][]{Guilloteau14}, inclined at 50\degr\ to the line of sight 
and featuring several density gaps suggesting ongoing planet formation \citep{Clarke18}.  More recently, interferometric K-band observations of the central regions of the accretion disc showed that CI~Tau also hosts an 
inner disc ring, whose inner edge is located at 0.2~au from the star, and that is misaligned with the main disc and tilted at 70\degr\ to the line of sight \citep{Gravity23}, presumably as a result of star-disc interactions.  
CI~Tau is indeed known to be strongly magnetic, with a large-scale field in excess of 1~kG \citep{Donati20b} and a small-scale field of about 2~kG \citep{Sokal20}.  A candidate eccentric close-in massive planet of minimum mass 
8.1~\mjup\ at an orbital period of 9~d was reported, on the basis of CI~Tau undergoing regular radial velocity (RV) variations with a semi-amplitude of about 1~\kms\ \citep{JohnsKrull16}, making this young star even more interesting 
to study.  However, spectropolarimetric observations of CI~Tau carried out with ESPaDOnS in the optical domain at the 3.6-m Canada-France-Hawaii Telescope (CFHT) unambiguously demonstrated that the 9-d period is actually 
the rotation period of the central star, and that the reported RV variations are in fact mostly due to stellar activity, and in particular to the presence of cool surface features coinciding with the footpoints of 
accretion funnels linking the star to the inner accretion disc, and carried in and out of the observer's view by rotation \citep{Donati20b}.  

To document this interesting case further and better characterize the immediate environment of CI~Tau, we carried out new observations, both with the near-infrared (nIR) SPIRou spectropolarimeter / velocimeter 
\citep{Donati20} and with its optical companion ESPaDOnS \citep{Donati03} at CFHT.  In this paper, we start by presenting the SPIRou and ESPaDOnS observations we collected (Sec.~\ref{sec:obs}), and briefly recall the main parameters 
of CI~Tau (Sec.~\ref{sec:par}).  We then analyse these new data, first measuring the longitudinal field \Bl\ (Sec.~\ref{sec:bl}) then carrying out a full tomographic imaging of the surface of CI~Tau (Sec.~\ref{sec:zdi}) 
in order to further constrain how magnetospheric accretion proceeds between the inner disc and the surface of CI~Tau (Sec.~\ref{sec:vei}).  We complete this study with a detailed RV analysis of whether CI~Tau 
may host close-in massive planets (Sec.~\ref{sec:rvs}) before summarizing our results and discussing their implications for our understanding of star / planet formation (Sec.~\ref{sec:dis}).

\section{SPIRou and ESPaDOnS observations}
\label{sec:obs}

CI~Tau was regularly observed from early 2019 to early 2023 with the SPIRou nIR spectropolarimeter / high-precision velocimeter \citep{Donati20} at CFHT, first within the SPIRou Legacy Survey (SLS) then within the 
SLS consolidation and enhancement programme SPICE.  Both the SLS and SPICE are large observational programmes (allocated 310 and 174~CFHT nights over 7 and 4 semesters respectively) dedicated to the study of 
planetary systems around nearby M dwarfs, and of star / planet formation in the presence of magnetic fields.  Additional observations of CI~Tau were collected in late 2020 with the ESPaDOnS optical spectropolarimeter 
\citep{Donati03} at CFHT, within the PI program of Jerome Bouvier (run ID 20BF01).  SPIRou and ESPaDOnS collect unpolarized and polarized stellar spectra, covering (in a single exposure) 
wavelength intervals of 0.95--2.50~\mic\ at a resolving power of 70\,000 for SPIRou, and of 0.37--1.00~\mic\ at a resolving power of 65\,000 for ESPaDOnS.  Each polarization sequence from both instruments consists
of 4 sub-exposures, associated with different orientations of the Fresnel rhomb retarders \cite[to remove systematics in polarization spectra to first order, see][]{Donati97b}, yielding one unpolarized (Stokes $I$), 
one circularly polarized (Stokes $V$) spectrum, and a null polarization check called $N$ allowing one to diagnose potential instrumental or data reduction problems.   

A total of 101 SPIRou polarization sequences of CI~Tau were collected in 3 main observing seasons, 34 from 2019 October to 2020 February, 39 from 2020 September to 2021 January and 28 from 2022 November to 2023 January, 
following an unsuccessful early start in 2019 February where 6 polarized sequences were recorded but discarded due to cold weather conditions in which the preliminary Fresnel rhombs used at this time did not behave 
properly retardation-wise\footnote{\emr The 2019 and 2020 SPIRou spectra of CI~Tau were also used by \citet{Sousa23} for a thorough study of infrared veiling of cTTSs.}.  
Poor weather conditions also led us to discard two polarized sequences in the first season, and another two in the last season, whereas an instrumental issue with the Fabry-Perot reference 
channel (strongly outshining the science channels) spoiled 3 polarization sequences in 2020 January.  This left us altogether with 94 usable Stokes $I$, $V$ and $N$ spectra of CI~Tau, with 29, 39 and 26 for the first, 
second and last season respectively, recorded over a total time span of almost 1200~d or 133 rotation cycles.  An additional 13 optical spectra of CI~Tau were collected with ESPaDOnS in 2020 November and December over a 
time frame of 16~d or 1.8 rotation cycles.  The full observation log for these spectra is given in Appendix~\ref{sec:appA}.    

Both SPIRou and ESPaDOnS data were processed with \texttt{Libre ESpRIT}, the nominal reduction pipeline of ESPaDOnS at CFHT, that was also adapted for SPIRou \citep{Donati20}.  
Least-Squares Deconvolution \citep[LSD,][]{Donati97b} was then applied to all reduced spectra, using a line mask constructed from the VALD-3 database \citep{Ryabchikova15} for an effective temperature \teff=4250~K and a 
logarithmic surface gravity \logg=4.0 adapted to CI~Tau (see Sec~\ref{sec:par}).  In these masks, we selected atomic lines of relative depth larger than 10 percent for SPIRou spectra (for a total of $\simeq$1500 lines, featuring 
an average wavelength and Land\'e factor of 1750~nm and 1.2 respectively), and of relative depth larger than 40 percent for ESPaDOnS spectra (for a total of $\simeq$9000 lines of average wavelength and Land\'e factor 640~nm 
and 1.2).  The line thresholds were set to ensure that the noise levels $\sigma_V$ in the resulting Stokes $V$ LSD profiles are similar in both sets of spectra, ranging from 1.9 to 6.2 (median 2.9, in units of 
$10^{-4} I_c$ where $I_c$ denotes the continuum intensity) 
for SPIRou spectra, and 2.6 to 4.0 (median 2.9) for ESPaDOnS spectra (in the same units).  In the case of SPIRou spectra, we also used a second mask restricted to CO lines between 2.2 and 2.4~\mic\ only and including 
in particular all CO lines of the 2.3~\mic\ CO bandhead (seen in absorption and of mostly stellar origin), which we use mainly in Sec.~\ref{sec:rvs}.  

All phases and rotation cycles associated with our SPIRou and ESPaDOnS spectra were computed assuming a rotation period of $\Prot=9.01$~d (see Sec.~\ref{sec:bl}) and counting from an arbitrary starting BJD0 of 2458762.0 (taken from 
our first usable SPIRou observation, see Table~\ref{tab:lgs}).

\section{Fundamental parameters of CI~Tau}
\label{sec:par}

We briefly recall here the main characteristics of CI~Tau, following our previous study based on older ESPaDOnS data \citep{Donati20b}, and complement it with new results from the refereed literature.  
older CI~Tau is a 2-Myr-old cTTS located at a distance of $160.3\pm0.4$~pc from the Sun \citep{Gaia23}, in the Taurus star forming region.  
CI~Tau has a mass of $0.90\pm0.02$~\msun\ \citep{Simon19}, a photospheric temperature of $\teff=4200\pm50$~K, a radius of $2.0\pm0.3$~\rsun, and rotates with a period of $\Prot=9.010\pm0.023$~d (see Sec.~\ref{sec:bl}).  This value 
of \Prot\ places the corotation radius \rcor, i.e., the distance from the centre of the star where the Keplerian orbital period in the equatorial plane equals \Prot, to $0.082\pm0.001$~au or 
equivalently $8.8\pm1.3$~\rstar.  
With a logarithmic luminosity relative to the Sun of $\log(\lstar/\lsun)=0.1\pm0.1$, CI~Tau is still in the process of completing its phase of nearly isothermal contraction along its Hayashi track, and is presumably 
still fully convective \citep{Donati20b}.  

Besides, CI~Tau is surrounded by a massive accretion disc inclined to the line-of-sight at an angle of $50\pm4$~\degr\ \citep{Guilloteau14} and extending up to a distance of several hundreds of au.  It features a succession of dusty 
rings and gaps located at radii of $\simeq$13, 39 and 100~au \citep{Clarke18} that suggests ongoing planet formation.  It was recently discovered that the inner regions of the accretion disc are significantly tilted 
with respect to the outer regions, with the inner disc being tilted at $70\pm1$\degr\ with respect to the line of sight \citep{Gravity23}.  The origin of the inner-outer disc misalignment is not clear, but could be due 
to magnetic warping caused by star-disc magnetic interactions \citep[e.g.,][]{Romanova21}, to exotic phenomena occurring within the disc \citep[e.g.,][]{Demidova23}, and / or to the presence of a  massive close-in planet embedded in the disc.  
In fact, such a planet was invoked for CI~Tau \citep{JohnsKrull16} but what was 
assumed to be the planet orbital period finally turns out to be the rotation period of the star, whereas the reported RV semi-amplitude (of $1.08\pm0.25$~\kms) thought to probe the reflex motion of the star under the 
gravitational pull of the planet is in fact attributable to stellar activity \citep{Donati20b}.  

We summarize in Table~\ref{tab:par} the main parameters of CI~Tau used in (or derived from) our study.  {\emr In particular, we assume in the following that the rotation axis of CI~Tau is inclined at $i=70$\degr\ to the line 
of sight, consistent with the adopted \vsini, \Prot, \teff\ and luminosity, and in line with the reported tilt angle of the inner disc. } 

\begin{table}
\caption[]{Parameters of CI~Tau used in / derived from our study} 
\scalebox{0.95}{\hspace{-4mm}
\begin{tabular}{ccc}
\hline
distance (pc)        & $160.3\pm0.4$   & \citet{Gaia23} \\
$A_V$ (mag)          & $0.65\pm0.20$   & \citet{Donati20b} \\ 
$\log(\lstar/\lsun)$ & $0.1\pm0.1$     & \citet{Donati20b} \\ 
\teff\ (K)           & $4200\pm50$     & \citet{Donati20b} \\
\mstar\ (\msun)      & $0.90\pm0.02$   & \citet{Simon19} \\
\rstar\ (\rsun)      & $2.0\pm0.3$     & \citet{Donati20b} \\
\logg\ (dex)         & $3.8\pm0.1$     & from mass and radius\\ 
age (Myr)            & $2\pm1$         & \citet{Donati20b} \\ 
\Prot\ (d)           & $9.01$          & period used to phase data \\ 
\Prot\ (d)           & $9.010\pm0.023$ & period from \Bl\ data \\ 
\vsini\ (\kms)       & $9.5\pm0.5$     & \citet{Donati20b} \\ 
$i$ (\degr), star           & $58_{-14}^{+32}$   & from \vsini, \Prot\ and \rstar \\ 
$i$ (\degr), star           & $70$      & used for ZDI \\ 
$i$ (\degr), inner disc     & $70\pm4$  & \citet{Gravity23} \\ 
$i$ (\degr), outer disc     & $50\pm4$  & \citet{Guilloteau14} \\ 
\rcor\ (au)          & $0.082\pm0.001$  & from \mstar\ and \Prot \\ 
\rcor\ (\rstar)      & $8.8\pm1.3$      &  \\ 
$\log\Mdot$ (\mspy)   & $-7.5\pm0.2$     & from \pab\ \& \brg, \citet{Sousa23} \\ 
$\log\Mdot$ (\mspy)   & $-8.5\pm0.2$     & from NC of \hei\ $D_3$ \& \caii\ IRT \\ 
\hline
\end{tabular}}
\label{tab:par}
\end{table}

\section{The longitudinal field of CI~Tau}
\label{sec:bl}

We pursued our analysis by computing the longitudinal field \Bl\ (i.e., the line-of-sight-projected component of the vector magnetic field averaged over the visible hemisphere) 
of CI~Tau for each of our observing epochs, from the Stokes $V$ and $I$ LSD profiles derived in Sec.~\ref{sec:obs}, and following \citet{Donati97b}.  The magnetic field of CI~Tau 
being rather strong and the nIR line profiles being significantly broadened by Zeeman broadening \citep{Sokal20}, Stokes $V$ LSD signatures were integrated over a window of $\pm30$~\kms\ 
(centred on the stellar rest frame) to estimate the first moment and its error bar, whereas the equivalent width of the Stokes $I$ LSD profiles is measured through a Gaussian fit.  
{\emr This window is large enough to ensure that no magnetic information is lost but small enough to minimize the resulting noise on \Bl, with small changes on the window size having minimal impact on the results. } 
As usual \citep[e.g.,][]{Donati97b, Donati23}, we also derive a pseudo longitudinal field from the polarization check $N$ in the exact same way we do 
from $V$, which allows us to check that it is consistent with 0 within the error bars, i.e., that it yields a reduced chi-square \chisqr\ close to 1.  The \Bl\ values corresponding to our 
94 validated SPIRou visits are listed in Table~\ref{tab:lgs}, and range from $-170$ to 120~G (median $-5$~G) with error bars of 10 to 30~G (median 15~G).  The corresponding \chisqr\ (with respect to $\Bl=0$~G line) 
is equal to 22, demonstrating that the large-scale field of CI~Tau is very clearly detected in the recorded Stokes $V$ Zeeman signatures;  the same operation applied to $N$ yields $\chisqr=0.92$, 
indicating that no spurious contamination is observed and that our formal error bars, dominated by photon noise, are consistent with the observed dispersion.  We note that our \Bl\ 
values are typically an order of magnitude weaker than the small-scale fields estimated from the Zeeman broadening of nIR lines \citep[$\simeq$2~kG,][]{Sokal20}, as usual for such active 
stars whose tangled small-scale fields generate circular polarization signatures that mostly cancel out and thereby do not contribute much to longitudinal fields \citep[e.g.,][]{Morin10,Donati23}.  

To investigate the quasi-periodic (QP) behaviour of \Bl\ and its evolution with time, we carry out a Gaussian-Process Regression (GPR) fit of the \Bl\ curve, with a covariance function $c(t,t')$ 
of the following type: 
\begin{eqnarray}
c(t,t') = \theta_1^2 \exp \left( -\frac{(t-t')^2}{2 \theta_3^2} -\frac{\sin^2 \left( \frac{\pi (t-t')}{\theta_2} \right)}{2 \theta_4^2} \right) 
\label{eq:covar}
\end{eqnarray}
where $\theta_1$ is the amplitude (in G) of the Gaussian Process (GP), $\theta_2$ its recurrence period (i.e., \Prot, in d), $\theta_3$ the evolution timescale (in d) on which the shape of 
the \Bl\ curve changes, and $\theta_4$ a smoothing parameter describing the amount of harmonic complexity needed to describe the data.  
We end up selecting the QP GPR fit to our \Bl\ points (arranged in a vector denoted $y$) that features the highest likelihood $\mathcal{L}$, defined by: 
\begin{eqnarray}
2 \log \mathcal{L} = -n \log(2\pi) - \log|C+\Sigma+S| - y^T (C+\Sigma+S)^{-1} y
\label{eq:llik}
\end{eqnarray}
where $C$ is the covariance matrix for our 94 epochs, $\Sigma$ the diagonal variance matrix associated with $y$ and $S=\theta_5^2 I$ the contribution from an additional white noise 
that we introduce as a fifth hyper-parameter $\theta_5$ (in case our error bars on \Bl\ were underestimated for some reason).  The hyper-parameter domain is then explored using a 
Monte-Carlo Markov Chain (MCMC) process, yielding posterior distributions and error bars for all hyper-parameters.  

\begin{table} 
\caption[]{Results of our MCMC modeling of the \Bl\ curve of CI~Tau with QP GPR.  For each hyper-parameter, we list the fitted value, the corresponding error bar and the assumed prior.  
The knee of the modified Jeffreys prior is set to $\sigma_{B}$, i.e., the median error bar of our \Bl\ measurements (i.e., 15~G). We also quote the resulting \chisqr\ and RMS 
of the final GPR fit. }  
\scalebox{0.9}{\hspace{-4mm}
\begin{tabular}{cccc}
\hline
Parameter   & Name & value & Prior   \\
\hline 
GP amplitude (G)     & $\theta_1$  & $77^{+20}_{-16}$  & mod Jeffreys ($\sigma_{B}$) \\
Rec.\ period (d)     & $\theta_2$  & $9.010\pm0.023$   & Gaussian (9.0, 3.0) \\
Evol.\ timescale (d) & $\theta_3$  & $146^{+42}_{-32}$ & log Gaussian ($\log$ 140, $\log$ 2) \\
Smoothing            & $\theta_4$  & $0.60\pm0.11$     & Uniform  (0, 3) \\
White noise (G)      & $\theta_5$  & $8^{+4}_{-3}$     & mod Jeffreys ($\sigma_{B}$) \\
\hline
\chisqr              &  \multicolumn{3}{c}{0.82}              \\ 
RMS (G)              &  \multicolumn{3}{c}{13.6}              \\ 
\hline 
\end{tabular}}      
\label{tab:gpr}      
\end{table}          

\begin{figure*}
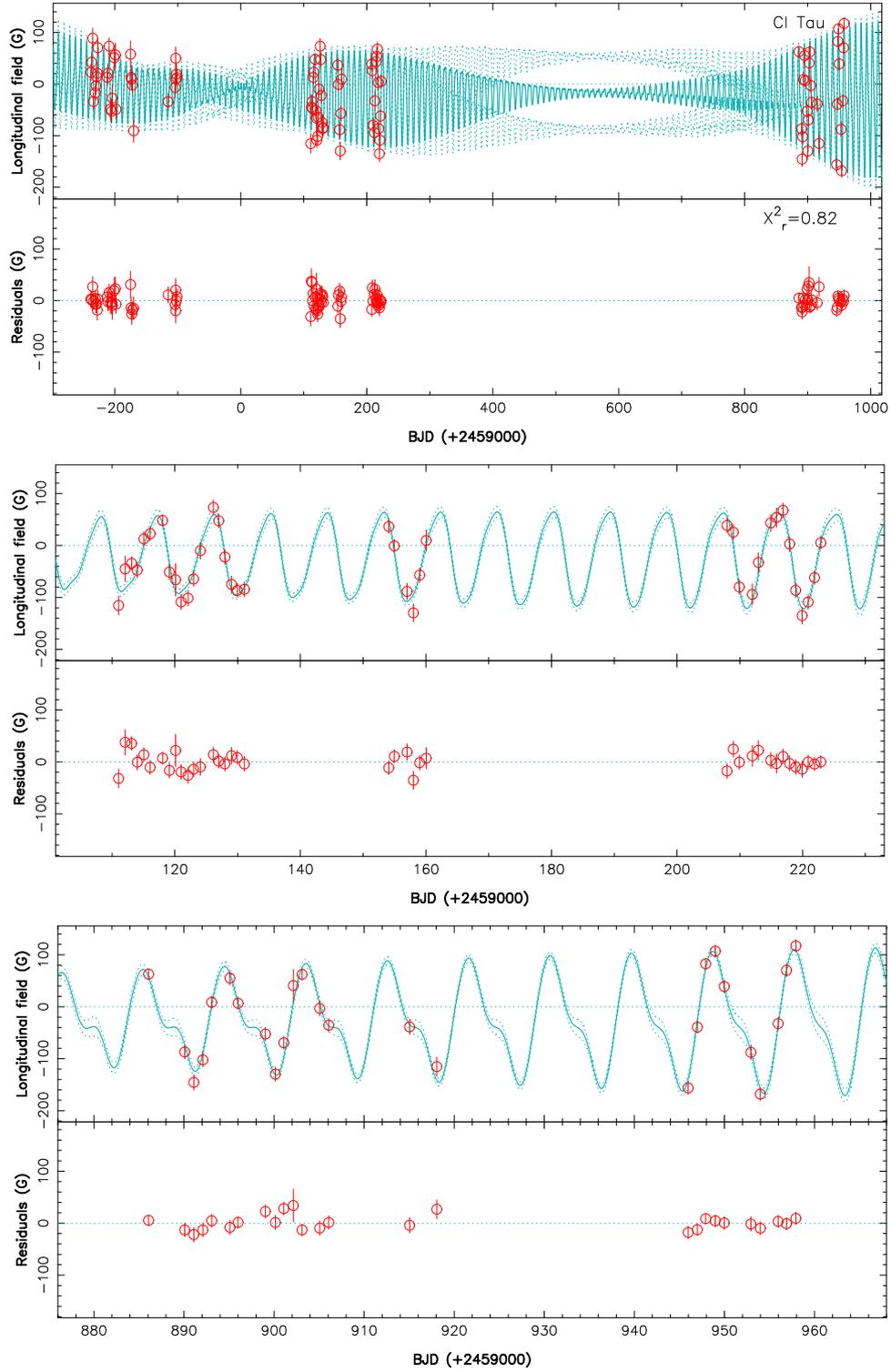

\centerline{\includegraphics[scale=0.49,angle=-90]{fig/citau2-gpb.ps} \vspace{2mm}}
\centerline{\includegraphics[scale=0.49,angle=-90]{fig/citau2-gpb20.ps} \vspace{2mm}}
\centerline{\includegraphics[scale=0.49,angle=-90]{fig/citau2-gpb22.ps}}
\caption[]{Longitudinal magnetic field \Bl\ of CI~Tau (red dots) as measured with SPIRou throughout our campaign, and QP GPR fit to the data (cyan full line) with corresponding 68\% confidence 
intervals (cyan dotted lines).  The residuals are shown in the bottom 
plot of each panel.  The top panel shows the whole data set, whereas the lower 2 panels present a zoom on the 2020 and 2022 data respectively. The RMS of the residuals is about 14~G, 
consistent with our median error bar of 15~G, yielding $\chisqr=0.82$ (whereas the \chisqr\ with respect to the $\Bl=0$~G line is 22).}  
\label{fig:gpb}
\end{figure*}

The results of the GPR fit are shown in Fig.~\ref{fig:gpb} (with a zoom on seasons 2020 and 2022 in the middle and lower panels), whereas the derived hyper-parameters are 
listed in Table~\ref{tab:gpr}.  The first conclusion is that $\Prot=9.010\pm0.023$~d, thereby confirming unambiguously the findings of our previous study \citep{Donati20b}.  
We stress in particular that our value of \Prot\ is the only one on which the MCMC systematically converged whatever prior we used and initial value we provided for this parameter.  
The measured period is in fact so well defined that the 1-yr aliases do not even show up in the associated corner plot.  
Besides, this means that the 6.6~d period quoted in several studies \citep[e.g.,][]{JohnsKrull16, Biddle18, Biddle21} can be safely rejected as a potential rotation period of CI~Tau.  

We also find that CI~Tau shows a \Bl\ curve with a rather simple, nearly sinusoidal and slowly-evolving modulation pattern, hence the relatively high value of the smoothing parameter 
($\theta_4=0.60\pm0.11$) and of the evolution timescale ($\theta_3=146^{+42}_{-32}$~d) compared to the more evolved young active star AU~Mic that exhibits a more complex and 
rapidly evolving \Bl\ curve \citep{Donati23}.  We finally outline that our \Bl\ values can be fitted within the noise level ($\chisqr=0.82$) and thus that the reconstructed additional 
white noise is consistent with zero.  

\begin{figure*}
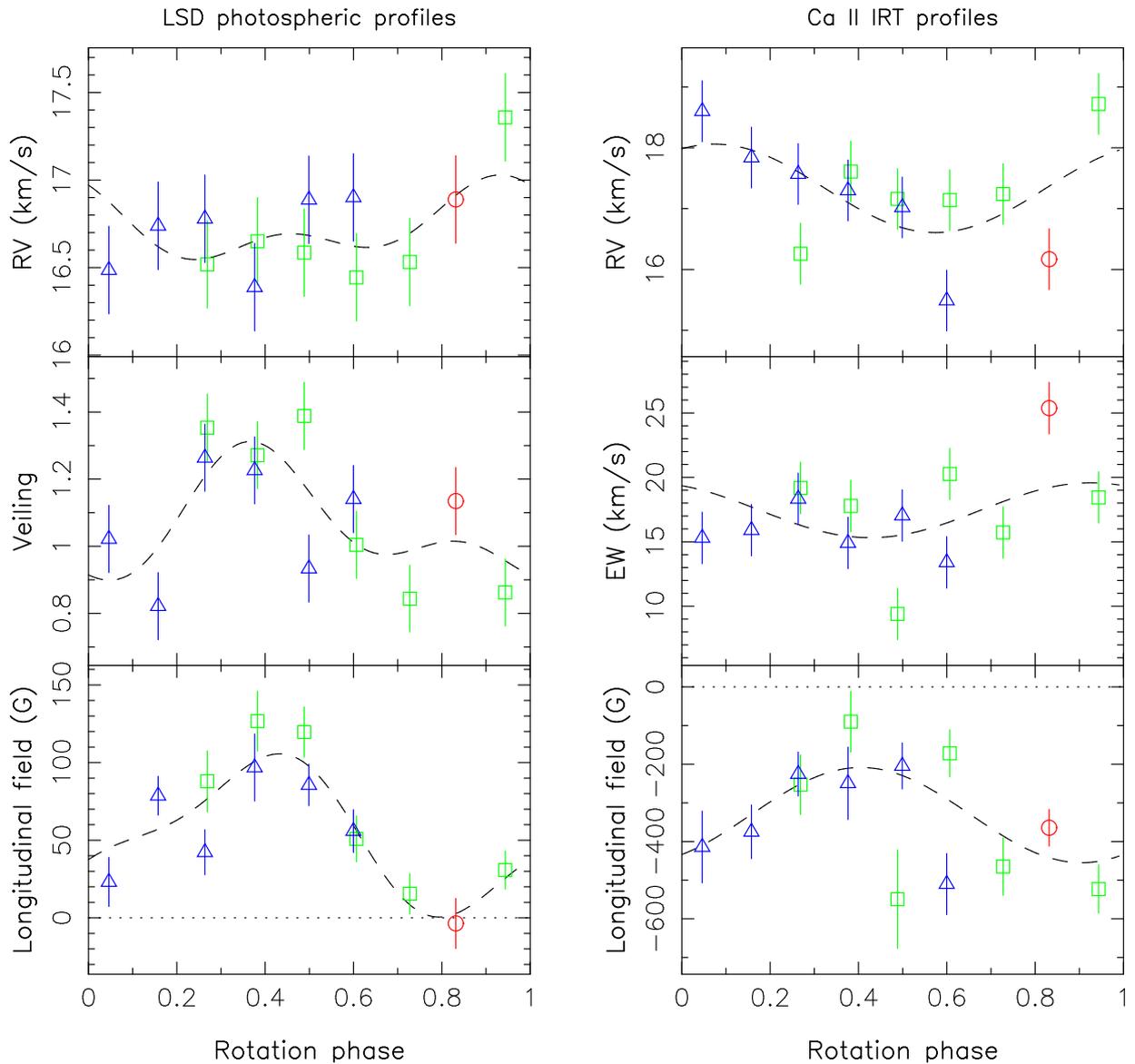

\centerline{\includegraphics[scale=0.6,angle=-90]{fig/citau2-pol.ps}\hspace{1cm}\includegraphics[scale=0.6,angle=-90]{fig/citau2-irt.ps}} 
\caption[]{Variability and modulation of the ESPaDOnS LSD profiles (left) and of the NC of the \caii\ IRT (right) lines, as a function of rotation 
phase (computed as mentioned in Sec.~\ref{sec:obs}).  Each panel shows the RVs (top), the veiled EWs (or veiling in the case of LSD profiles, middle) and the \Bl\ values (bottom).
The red circle, green squares and blue triangles depict measurements obtained during rotation cycles 45, 46 and 47 (see Table~\ref{tab:lge}).  {\emr The dashed lines are sine fits 
to the phase-folded data points, including the first harmonic in the left column and the fundamental only in the right one.} }   
\label{fig:esp}
\end{figure*}

We proceeded in the same way to derive \Bl\ values from our ESPaDOnS LSD profiles of CI~Tau from late 2020, now integrating over a smaller window of width $\pm20$~\kms\ (in the stellar rest frame) 
given that Stokes $I$ and $V$ LSD profiles derived from the optical domain are significantly narrower than those inferred at nIR wavelengths.  The 13 \Bl\ values we obtained from 
photospheric lines range from $-5$ to 130~G (median 55~G) with error bars of 12 to 22~G (median 15~G), and are shown in Fig.~\ref{fig:esp} (left panel).  We can already note that the corresponding 
\Bl\ curve is significantly different from that obtained in our previous study \citep[with data from early 2017,][]{Donati20b}, which showed two clear minima over a rotation period whereas that from our 2020 data only 
shows one.  

We also derived \Bl\ values from the 
narrow emission component (NC) of the \caii\ infrared triplet (IRT) lines (see Fig.~\ref{fig:esp}, right panel) and of the \hei\ D$_3$ line (not shown), both known to probe the post-shock region 
at the footpoints of accretion funnels (in addition to the quieter, non-accreting chromosphere for the \caii\ IRT lines), integrating over a wide window ($\pm30$ and $\pm50$~\kms\ for the \caii\ IRT 
and \hei\ D$_3$ line respectively) to take into account the large intrinsic widths of these 
lines\footnote{To isolate the NC from the broad emission component of the \caii\ IRT and \hei\ D$_3$ line, we proceed as in our previous studies, by fitting all profiles with a 
linear combination of broad Lorentzians and one narrow Gaussian functions, the first ones to match the broad emission component and the second one to adjust the NC.  Subtracting the fitted 
broad component from the observed profile gives access to the NC.  The clear difference in width between both components renders this separation relatively easy to achieve, unambiguous and reliable.}.    

A sine fit (plus first harmonic) to the \Bl\ points from ESPaDOnS LSD profiles of photospheric lines yields a period of $9.0\pm0.3$~d (see Fig.~\ref{fig:esp}, bottom left panel), fully consistent (though expectedly 
less accurate) than that derived from our SPIRou data (spanning a much longer time frame of 1200~d instead of only 16~d).  Doing the same on the \Bl\ data derived from the NC of the \caii\ 
IRT lines is less successful, the amplitude of the modulation pattern being less clear as a result of intrinsic variability affecting more strongly accretion lines than photospheric lines 
(see Fig.~\ref{fig:esp}, bottom right panel);  we nonetheless see a modulated trend, with \Bl\ being stronger and more negative in the second half of the rotation cycle.  \Bl\ values 
derived from the \hei\ D$_3$ line show no clear modulation (within error bars of $\simeq$300~G), and stay more or less constant at $-1$~kG.  This is again different from the results of 
our previous study where the corresponding \Bl\ curve displayed a very obvious modulation, varying from 0 to --2~kG in a more or less sinusoidal fashion \citep{Donati20b}.  

One can note (see, e.g., Fig.~\ref{fig:esp} and Table~\ref{tab:lge}) that \Bl\ values from ESPaDOnS LSD profiles are discrepant with those from SPIRou LSD profiles collected a few weeks before and after the ESPaDOnS observations, 
the SPIRou \Bl\ curve reaching down to $-130$~G whereas the ESPaDOnS \Bl\ points are almost all positive.  Similarly, \Bl\ values from the \caii\ IRT and \hei\ D$_3$ lines are constantly negative, 
unambiguously indicating that the accreting regions on the visible hemisphere at the surface of the star are associated with magnetic patches of negative polarity.  Despite this discrepancy, 
we can see that the epoch at which \Bl\ reaches its minimum value when measured from ESPaDOnS LSD profiles ($-3.6\pm16.1$~G, on 2020 November 20 at BJD~$\simeq$~2459175.94, see Table~\ref{tab:lge}) 
coincides well with that at which \Bl\ is expected to reach its strongest negative value at SPIRou wavelengths (see Fig.~\ref{fig:gpb} middle panel);  this is also the time at which \Bl\ from \caii\ 
IRT lines is most negative (see Fig.~\ref{fig:esp}, bottom right panel).  Altogether, it implies that all three \Bl\ curves are in phase though different in amplitude, shape and vertical offset.  

These apparent inconsistencies can in fact be exploited to our advantage in order to more finely characterize what happens at the surface of CI~Tau.  
We propose they directly reflect (i) that photospheric brightness is non-uniform at the surface of CI~Tau, (ii) that accreting regions of negative polarity (bright at chromospheric level) tend to coincide 
with cool regions at photospheric level, that are darker in the optical than in the nIR (as a result of the wavelength dependence of the spot-to-photosphere contrast), and (iii) these dark regions 
face the observer around phase $\simeq$0.8 in our ephemeris (see Sec.~\ref{sec:obs}) and are located at high latitudes on the visible hemisphere given the low amplitude modulation ($<$1~\kms) of 
the RV curves from both emission line NCs (see, e.g., Fig.~\ref{fig:esp} top right panel for the \caii\ IRT).  
This preliminary conclusion is qualitatively consistent with the findings of our previous work on CI~Tau \citep[][]{Donati20b} as well as of similar spectropolarimetric studies of cTTSs 
\citep[e.g.,][]{Donati11}.  Our combined SPIRou and ESPaDOnS data set of CI~Tau from 2020 September to 2021 January is thus quite unique in this respect, yielding an unprecedented richness of 
complementary diagnostics on the large-scale magnetic field, on the photospheric brightness and on the ongoing embedded accretion flows.  
We elaborate along these lines in a more quantitative way in Sec.~\ref{sec:zd2}, by simultaneously modelling the Stokes $I$ and $V$ profiles of our three sets of lines (SPIRou LSD profiles, 
ESPaDOnS LSD profiles, the NC of the \caii\ IRT) for this specific observing season.

\section{Zeeman-Doppler Imaging of CI~Tau}
\label{sec:zdi}

From time series of Stokes $V$ and $I$ LSD profiles, one can reconstruct the magnetic field and brightness distribution at the surface of an active star like CI~Tau using Zeeman-Doppler Imaging (ZDI), a tomographic 
technique inspired from medical imaging that allows one to invert phase-resolved sets of line profiles into maps of the large-scale vector magnetic field and of the photospheric brightness at the 
surface of the star.  

\subsection{Zeeman-Doppler Imaging}
\label{sec:zd0}

To achieve this goal, ZDI proceeds iteratively, starting from a small magnetic seed and a featureless brightness map, and progressively building up information by exploring the parameter space using a variant 
of the conjugate gradient method \citep[e.g.,][]{Skilling84, Brown91, Donati97c, Donati06b, Kochukhov16}.  At each iteration, ZDI adds features on the maps by comparing the observed Stokes profiles with 
the synthetic ones associated with the current images.  The aim is to obtain a given agreement with the data (i.e., a given \chisqr, usually 1) with minimal information in the maps to ensure that 
whatever is reconstructed is mandatorily required by the data.  This reflects the fact that the problem is ill-posed and allows an infinite number of solutions of variable complexity, among which ZDI 
selects the simplest one that matches the data at the requested level of \chisqr.  When optical data are also available (as for our 2020 observing season), one can constrain the large-scale 
field and brightness map by simultaneously using Stokes $I$ and $V$ LSD profiles of photospheric lines from both domains, explicitly taking into account that the spot-to-photosphere brightness contrast  
differs at optical and nIR wavelengths.  {\emr Furthermore, one can also reconstruct at the same time with ZDI a surface map of the accretion-induced excess emission in specific emission lines, the NC of the 
\caii\ IRT in our case, from time-series of their profiles \citep[e.g.,][]{Donati10b, Donati11};  such maps inform us on where, at the surface of the star, the local line emission, boosted by accretion, 
gets much larger than the quiet chromospheric emission. }  

In practice, we start by dividing the surface of the star into a grid of a few thousand (typically 5000) cells so that synthetic Stokes $I$ and $V$ profiles at each observation epoch can be computed by 
summing up the contributions of all cells in the reconstructed images, taking into account the assumed stellar parameters, in particular the inclination of the rotation axis to the line of sight $i$ 
(set to 70\degr, see Table~\ref{tab:par}), the line-of-sight-projected equatorial velocity at the surface of the star \vsini\ \citep[set to 9.5$\pm$0.5~\kms\ following][]{Donati20b}, and the linear limb 
darkening coefficient (set to 0.3 and 0.7 at nIR and optical wavelengths respectively).  We also assume that the surface of CI~Tau rotates as a solid body, consistent with previous findings 
that fully-convective TTSs similar to CI~Tau show weak levels of surface differential rotation \citep[e.g.,][]{Finociety23}.  

Local Stokes $I$ and $V$ contributions from each cell are derived using Unno-Rachkovsky's analytical 
solution of the polarized radiative transfer equation in a plane-parallel Milne Eddington atmosphere \citep{Landi04}, assuming a Land\'e factor and average wavelength of 1.2 and 1750~nm for the SPIRou 
photospheric LSD profiles, and 1.2 and 640~nm for the ESPaDOnS ones.  For the Doppler width \vD\ of the local profile, we assume $\vD=3$~\kms, both for SPIRou and ESPaDOnS data, a value that yields 
synthetic LSD profiles consistent with those of CI~Tau and other slowly rotating TTSs.  {\emr The brightness maps that ZDI recovers consist of sets of independent pixels describing the local brightness 
(relative to the quiet photosphere and denoted $b$) over all surface grid cells.}   When using SPIRou and ESPaDOnS data simultaneously, $b$ is assumed to depend on wavelength in a way where $b_{\rm S}$ at 
1750~nm can be simply inferred from $b_{\rm E}$ at 640~nm by $b_{\rm S} = {b_{\rm E}}^{\delta}$.  If the spot brightness obeys Planck's law, we should have $\delta\simeq0.40$;  if however the spot contrast 
depends much more steeply on wavelength \citep[as for, e.g., the wTTS V410~Tau,][]{Finociety21}, $\delta$ can end up being much smaller.  

In the case of the \caii\ IRT profiles, we proceed as in previous similar papers \citep[e.g.,][]{Donati10b,Donati11,Donati12}, i.e., by applying to the data a filtering procedure whose aim is to retain rotational 
modulation only and discard intrinsic variations, hence helping the convergence of the imaging code.  The local profile we use to describe this line has a wavelength 850~nm, a Doppler width $\vD=7$~\kms\ 
(ensuring synthetic profiles consistent with those of CI~Tau) and a Land\'e factor 1.0.  {\emr The quantity ZDI recovers, again handled as a set of independent pixels, describes the fractional area of each surface pixel 
in which accretion occurs, and where the local line emission flux, boosted by accretion, is assumed to be larger than that in the quiet surrounding chromosphere by an accretion-induced emission enhancement factor 
$\epsilon$.  As in our previous papers \citep[e.g.,][]{Donati10b, Donati11, Donati12}, we set $\epsilon=10$ for the current study, with the exact value of $\epsilon$ having no more than a global scaling 
effect on the recovered images (ensuring that the integrated fractional accretion area times $\epsilon$ is constant). } 

Last but not least, the magnetic field at the surface of the star is described as a spherical harmonics (SH) expansion, using the formalism of \citet{Donati06b} in which the poloidal and toroidal components 
of the vector field are expressed with 3 sets of complex SH coefficients, $\alpha_{\ell,m}$ and $\beta_{\ell,m}$ for the poloidal component, and $\gamma_{\ell,m}$ for the toroidal component\footnote{In this 
paper, we use modified and more consistent expressions for the field components, where $\beta_{\ell,m}$ is replaced with $\alpha_{\ell,m}+\beta_{\ell,m}$ in the equations of the meridional and azimuthal field 
components \citep[see, e.g.,][]{Lehmann22, Finociety22,Donati23}.}, where $\ell$ and $m$ note the degree and order of the corresponding SH term in the expansion.  In the case of CI~Tau and given the low 
\vsini, we only use SH terms up to  $\ell=5$, which in practice is enough for describing the relatively simple large-scale field topology that the star hosts.  We further assume that the field is mostly 
antisymmetric with respect to the centre of the star \citep[i.e. that odd SH modes dominate, as in, e.g.,][]{Donati11}, so that accretion funnels linking the inner disc to the star are anchored at high 
latitudes, in agreement with our observations (see Sec.~\ref{sec:bl}).  In practice, this is achieved by penalizing even SH modes with respect to odd ones in the entropy function.  
Finally, we assume that only a fraction $f_V$ of each grid cell (called filling factor of the large-scale field, equal for all cells) actually contributes to Stokes $V$ profiles, with a magnetic flux 
over the cell equal to $B_V$ (and thus a magnetic field within the magnetic portion of the cells equal to $B_V/f_V$).  Similarly, we assume that a fraction $f_I$ of each grid cell (called filling factor 
of the small-scale field, again equal for all cells) hosts small-scale fields of strength $B_V/f_V$ (i.e., with a small-scale magnetic flux over the cell equal to $B_I = B_V f_I/f_V$).  This simple model 
implies in particular that the small-scale field locally scales up with the large-scale field (with a scaling factor of $f_I/f_V$), which is likely no more than a rough approximation, but at least ensures 
that the resulting Zeeman broadening from small-scale fields is consistent with the reconstructed large-scale field.  For our study, we set $f_I\simeq0.8$ for all lines, consistent with actual 
measurements \citep{Sokal20}, whereas we set $f_V\simeq0.4$ (again for all lines), slightly larger though still consistent with the value used in our previous study \citep[i.e., 0.3,][]{Donati20b} and 
yielding optimal fits to the data.

\subsection{ZDI from SPIRou data only} 
\label{sec:zd1}

\begin{figure*}
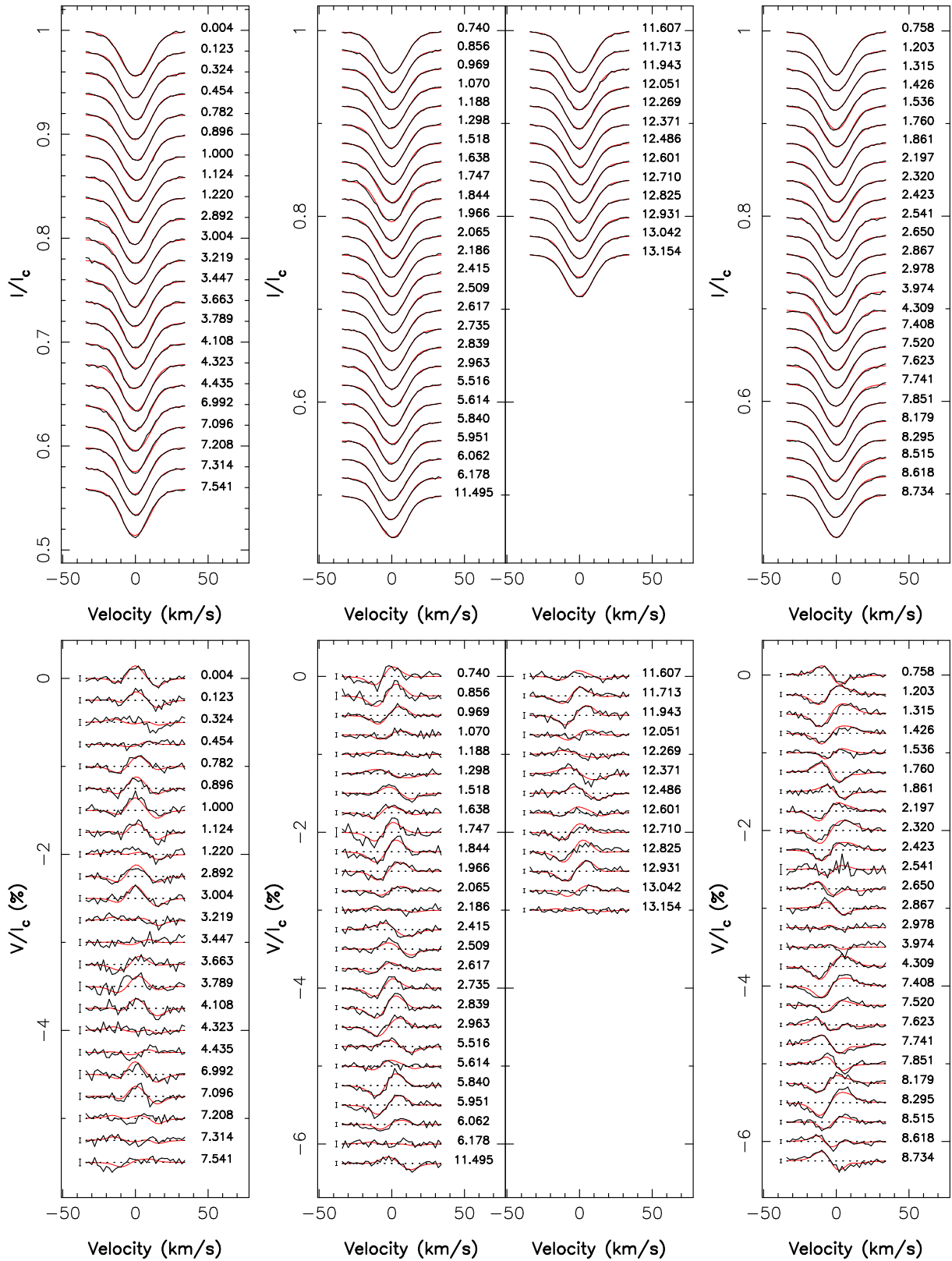

\centerline{\includegraphics[scale=0.6,angle=-90]{fig/citau2-lsdi19.ps}\hspace{2mm}\includegraphics[scale=0.6,angle=-90]{fig/citau2-lsdi20.ps}\hspace{2mm}\includegraphics[scale=0.6,angle=-90]{fig/citau2-lsdi22.ps}\vspace{2mm}} 
\centerline{\includegraphics[scale=0.6,angle=-90]{fig/citau2-lsdv19.ps}\hspace{2mm}\includegraphics[scale=0.6,angle=-90]{fig/citau2-lsdv20.ps}\hspace{2mm}\includegraphics[scale=0.6,angle=-90]{fig/citau2-lsdv22.ps}} 
\caption[]{Observed (thick black line) and modelled (thin red line) LSD Stokes $I$ (top panels) and $V$ (bottom) profiles of CI~Tau from SPIRou data, for seasons 2019 (left), 2020 (middle) and 2022 (right).  Rotation cycles (counting 
from 0, 38 and 124 for the 2019, 2020 and 2022 seasons respectively, see Table~\ref{tab:lgs}) are indicated to the right of all LSD profiles, while $\pm$1$\sigma$ error bars are added to the left of Stokes $V$ signatures. } 
\label{fig:fit}
\end{figure*}

\begin{figure*}
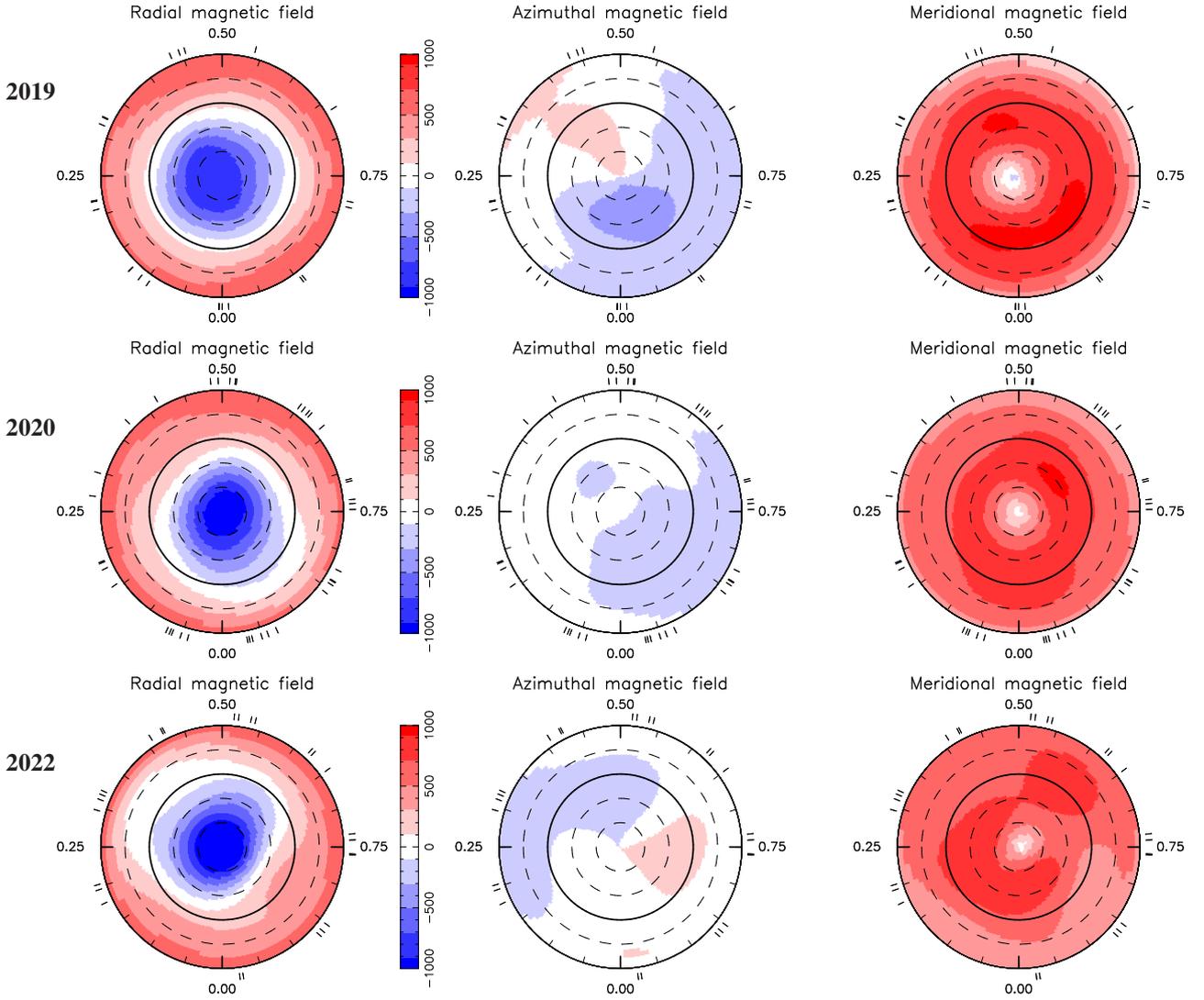

\centerline{\large\bf 2019\raisebox{0.3\totalheight}{\includegraphics[scale=0.45,angle=-90]{fig/citau2-map19.ps}}\vspace{2mm}}
\centerline{\large\bf 2020\raisebox{0.3\totalheight}{\includegraphics[scale=0.45,angle=-90]{fig/citau2-map20.ps}}\vspace{2mm}}
\centerline{\large\bf 2022\raisebox{0.3\totalheight}{\includegraphics[scale=0.45,angle=-90]{fig/citau2-map22.ps}}}
\caption[]{Reconstructed maps of the large-scale field of CI~Tau (left, middle and right columns for the radial, azimuthal and meridional components in spherical coordinates, in G), for seasons 2019, 2020 and 2022 (top to bottom row 
respectively), derived with ZDI from the SPIRou Stokes $I$ and $V$ LSD profiles of Fig.~\ref{fig:fit}.  The maps are shown in a flattened polar projection down to latitude $-60$\degr, with the north pole at the centre and the equator 
depicted as a bold line.  Outer ticks indicate phases of observations.  Positive radial, azimuthal and meridional fields respectively point outwards, counterclockwise and polewards. } 
\label{fig:map}
\end{figure*}

\begin{table*} 
\caption[]{Properties of the large-scale and small-scale magnetic field of CI~Tau for our 3 observing seasons, using our SPIRou data only (columns 2 to 6) or SPIRou and ESPaDOnS data simultaneously (in 2020 only, columns 7 to 11).  
We list the average reconstructed large-scale field strength <$B_V$> (columns 2 and 7), the maximum small-scale field strength $B_I$ (columns 3 and 8), the polar strength of the dipole component \Bd\ (columns 4 and 9), the tilt / phase of the 
dipole field to the rotation axis (columns 5 and 10) and the amount of magnetic energy reconstructed in the poloidal component of the field and in the axisymmetric modes of this component (columns 6 and 11).  Error bars 
are typically equal to 5--10\% for field strengths and percentages, and 5--10\degr\ for field inclinations. } 
\center{\scalebox{1.0}{\hspace{0mm}
\begin{tabular}{cccccccccccc}
\hline
       & \multicolumn{5}{c}{SPIRou data only}                        && \multicolumn{5}{c}{SPIRou \& ESPaDOnS data}    \\
\hline
Season & <$B_V$> & max $B_I$ & \Bd     & tilt / phase    & poloidal / axisym && <$B_V$> & max $B_I$ & \Bd     & tilt / phase   & poloidal / axisym \\ 
       &  (G)    &   (kG)    &  (G)    & (\degr) & (\%)              &&  (G)    &   (kG)    &  (G)    & (\degr) & (\%)              \\ 
\hline
2019   & 860 & 1.8 & 780 & 11 / 0.1 & 98 / 96 &&     &     &      &          &         \\ 
2020   & 820 & 2.1 & 790 & 18 / 0.9 & 99 / 97 && 840 & 2.4 & 1080 & 15 / 0.9 & 99 / 97 \\  
2022   & 810 & 2.6 & 820 & 18 / 0.4 & 99 / 94 &&     &     &      &          &         \\
\hline 
\end{tabular}}}
\label{tab:mag}
\end{table*}

We start by applying ZDI to our SPIRou LSD profiles of CI~Tau only, for each of our 3 observing seasons, i.e., 2019 (2019 October to December), 2020 (2020 September to 2021 January) and 2022 (2022 November to 2023 January).  
Note that the 6 profiles recorded in 2020 February (see Table~\ref{tab:lge}) were left out of the 2019 season set, as they were collected 2 months after the bulk of the other 23 profiles, at a time where 
the large-scale field already started to evolve significantly according to the \Bl\ curve (see Fig.~\ref{fig:gpb} top panel).  The resulting ZDI fits to the LSD Stokes $I$ and $V$ profiles are shown in Fig.~\ref{fig:fit} 
whereas the resulting maps of the large-scale field are presented in Fig.~\ref{fig:map}.  Stokes $V$ profiles are fitted at an average \chisqr\ level of about 1.2 for all seasons, i.e. slightly larger than 1 as a likely result 
of the field being subject to intrinsic variability over each observing season, both on short (e.g., accretion) and medium (e.g., convection) timescales (see, e.g., Fig.~\ref{fig:gpb}).  

Although brightness maps are reconstructed at the same time as magnetic maps when ZDI simultaneously fits Stokes $I$ and $V$ data, 
we find that CI~Tau exhibits no obvious brightness surface feature that is large enough and / or with a high enough contrast in the nIR with respect to the surrounding photosphere to generate clear Stokes $I$ profile distortions 
(see Fig.~\ref{fig:fit}) and thereby to show up in our reconstructed brightness images (not shown).  
This is in contrast with the brightness images in the optical domain that we reconstructed in our previous study, that showed high-contrast cool spots associated with strong magnetic field regions \citep{Donati20}.  

From the reconstructed magnetic maps, we conclude that CI~Tau hosts a large-scale magnetic field of average strength $\simeq$0.8--0.9~kG over the star at all times, reaching a maximum intensity of 
0.9, 1.1 and 1.3~kG in seasons 2019, 2020 and 2022.  Taking into account the assumed scaling factor $f_I/f_V\simeq2$ between the small-scale and large-scale field (see Sec.~\ref{sec:zd0}), this implies a small-scale field of order 
1.6--1.8~kG, reaching up to 2.6~kG, in reasonable agreement with literature results \citep{Sokal20}.  The large-scale field  we reconstruct is almost fully poloidal and axisymmetric, and mainly consists of a 0.8~kG dipole inclined 
at 11\degr\ (in 2019) to 18\degr\ (in 2020 and 2022) to the rotation axis, towards phases 0.1, 0.9 and 0.4 in seasons 2019, 2020 and 2022 respectively.  The quadrupole and octupole components are weak, at most 0.3~kG for the latter 
in 2022.  We summarize the magnetic properties of the reconstructed magnetic topologies in Table~\ref{tab:mag}.  

What it suggests already is that the large-scale field of CI~Tau significantly changed since our previous study \citep{Donati20b}, with the dipole field being weaker by potentially up to a factor of 2 and the octupole component having 
almost vanished (weaker by an order of magnitude).  We cannot entirely rule out that part of this difference reflects the different data used in both study \citep[SPIRou nIR data here and ESPaDOnS optical data 
in][]{Donati20b}.  However, looking at the new set of ESPaDOnS data we collected in season 2020 and in particular the \Bl\ curves (see Sec.~\ref{sec:bl} and Fig.~\ref{fig:esp}), we can already confirm that significant changes occurred 
in the large-scale of CI~Tau (see Sec.~\ref{sec:bl}).  The following section gives further support in this direction.

\begin{figure*}
\centerline{\large\bf 2020\raisebox{0.1\totalheight}{\includegraphics[scale=0.68,angle=-90]{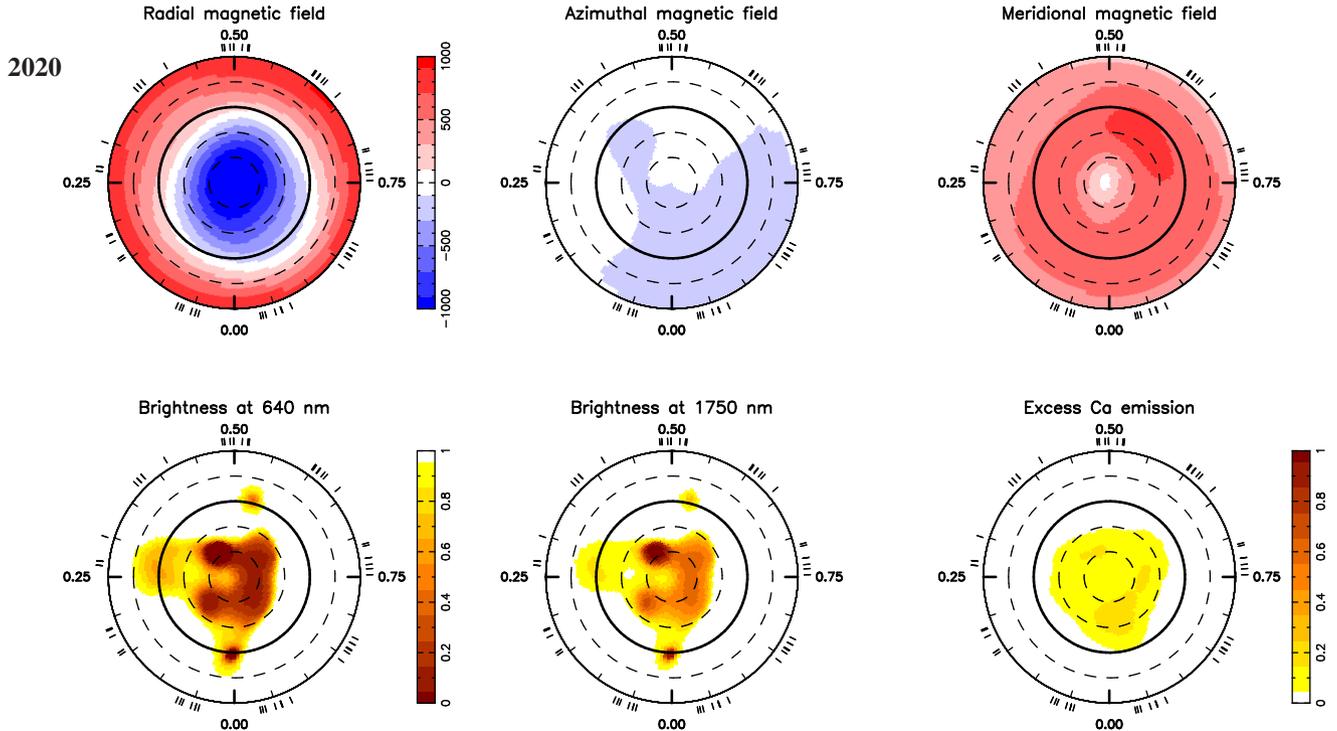}}}
\caption[]{Same as Fig.~\ref{fig:map} for the 2020 season, using now all three sets of lines from ESPaDOnS and SPIRou data and assuming $\delta=0.4$ (see Sec.~\ref{sec:zd2}).  Below the three components of the magnetic field (top row), 
we show the reconstructed brightness map as seen at ESPaDOnS wavelengths (left) and SPIRou wavelengths (middle), along with the map of accretion-induced excess emission in the NC of the \caii\ IRT (bottom right).  
Darker color levels indicate darker photospheric features for the bottom left and middle plots, and brighter chromospheric features for the bottom right plot.  } 
\label{fig:mapes}
\end{figure*}

\subsection{ZDI from SPIRou and ESPaDOnS data} 
\label{sec:zd2}

For our 2020 data set, we also applied ZDI to our combined SPIRou and ESPaDOnS data, using now three independent sets of lines (LSD photospheric profiles from both SPIRou and ESPaDOnS spectra, and the \caii\ IRT from the ESPaDOnS spectra) 
to characterize in a more self-consistent way the large-scale and small-scale field of CI~Tau in this specific season.  As pointed out in Sec.~\ref{sec:bl}, the longitudinal field curves associated with our three sets of lines, despite 
varying in phase, are apparently inconsistent, with ESPaDOnS photospheric lines showing mainly positive longitudinal fields, the NC of \caii\ IRT lines exhibiting negative fields only, and SPIRou photospheric lines displaying both positive 
and negative longitudinal fields (see Figs.~\ref{fig:gpb} and \ref{fig:esp}).  The simplest qualitative explanation for this apparent discrepancy is that dark spots are present at the surface of the star at ESPaDOnS wavelengths, obscuring 
some of the most magnetic regions and preventing them from significantly contributing to Stokes $V$ signatures at optical wavelengths.  However, as we have seen in Sec.~\ref{sec:zd1}, brightness images reconstructed simultaneously with magnetic 
maps from SPIRou data show virtually no feature at the surface of the star, either because these features have a very low contrast with respect to the photosphere at SPIRou wavelengths, and / or because they are minor contributions to the 
line profile distortion (with respect to, e.g., magnetic fields).  We explore both options below.  

We start by assuming that spot brightness varies with wavelength as expected from Planck's law, and thus that $\delta$, the exponent with which relative brightness at SPIRou wavelengths $b_{\rm S}$ can be inferred from that at ESPaDOnS 
wavelengths $b_{\rm E}$, is equal to $\delta\simeq0.4$ (see Sec.~\ref{sec:zd0}).  In this context, we are able to obtain a convincing fit to all three sets of lines (see Fig.~\ref{fig:fite}).  The corresponding ZDI maps are shown in 
Fig.~\ref{fig:mapes}, where we present the reconstructed brightness map at both ESPaDOnS and SPIRou wavelengths, along with the map of accretion-induced excess emission in the NC of the \caii\ IRT lines.  We obtain 
that the reconstructed large-scale field reaches an averaged strength of 840~G and a maximum intensity of 1.2~kG, i.e., slightly larger than the values obtained from SPIRou data only (see Sec.~\ref{sec:zd1}).  
The derived magnetic topology is again mostly poloidal and axisymmetric, as previously indicated by SPIRou data.  The dipole component of the large-scale field reaches almost 1.1~kG, i.e., $\simeq$35 percent larger than our estimate 
from SPIRou data alone (thanks to the additional constraints from the Stokes $I$ and $V$ ESPaDOnS data) and is tilted at $\simeq$15\degr\ to the rotation axis towards phase 0.90.  We 
also confirm that the quadrupole and octupole components of the large-scale field are weak, smaller than 150~G in season 2020.  

The map of accretion-induced excess emission in the NC of the \caii\ IRT lines (see Fig.~\ref{fig:mapes}, bottom right) demonstrates that accretion occurs mostly towards a region that is centred on the pole and extends to intermediate 
latitudes, {\emr consistent with a low-amplitude modulation of emission peaking around phase 0.9 (see Fig.~\ref{fig:esp} middle right panel)}, i.e., the phase towards which the dipole component of the reconstructed field is tilted.  
It is also the phase at which the longitudinal field as measured in SPIRou LSD profiles of 
photospheric lines (see Fig.~\ref{fig:gpb} and Table~\ref{tab:lge}) and in the NC of the \caii\ IRT lines (see Fig.~\ref{fig:esp} and Table~\ref{tab:lge}) is strongest, at least when \caii\ IRT \Bl\ values are averaged over the two rotation cycles we monitored with 
ESPaDOnS.  This is consistent with the footpoints of accretion funnels linking the stellar 
surface to the inner accretion disc being anchored at high latitudes on CI~Tau, close to the region of maximum magnetic field.  It also matches the longitudinal field measured in the NC of the \hei\ D$_3$ line ($\simeq$--1~kG, 
see Sec.~\ref{sec:bl}), equal to the projection of the field vector in the accretion region where the average field is inclined at $\simeq$55\degr\ to the line of sight.  

The reconstructed brightness image of CI~Tau shown in Fig.~\ref{fig:mapes} is predicted to generate photometric variations with a full amplitude of 4.5\% in the ESPaDOnS domain ($I$ band), and 1.8\% in the SPIRou domain ($H$ band), both 
of which would be hardly detectable amidst accretion-induced photometric perturbations causing intrinsic photometric fluctuations at a level of $\simeq$0.1~mag (i.e., 10\%) RMS for CI~Tau (see Fig.~\ref{fig:esp} middle left panel, and 
Manick et al, submitted).  It  
explains why the rotation period is hard to detect from a simple light-curve periodogram, especially in seasons where the large-scale magnetic topology, and thereby the brightness distribution at the surface of the star, is 
close to being axisymmetric, as was the case in our 2020 observing season.  The derived maps are also consistent, by construction, with the observed RV variations in the various spectral lines used for the ZDI imaging, that do not exceed 
a peak-to-peak amplitude of 0.5~\kms\ for photospheric lines and 1~\kms\ for the NC of the \caii\ IRT (see, e.g., Fig.~\ref{fig:esp}).  We come back in Sec.~\ref{sec:rvs} for a more extensive discussion on the SPIRou RVs of CI~Tau.  
As a sanity check, we also verified that the maps reconstructed from our 2020 ESPaDOnS data alone (using both Stokes $I$ and $V$ LSD profiles of photospheric lines and those of the NC of the \caii\ IRT) are consistent with those using both 
SPIRou and ESPaDOnS data (shown in Fig.~\ref{fig:mapes}).  

Assuming now that $\delta$ is much smaller than 0.4 to ensure that the reconstructed brightness map at 1750~nm is as featureless as those derived in Sec.~\ref{sec:zd1} from SPIRou data alone, we find in practice that $\delta$ needs to be as 
small as 0.01 to obtain this result.  This would mean in particular that brightness features at the surface of CI~Tau at SPIRou wavelengths are drastically less contrasted than, and cannot be simply derived with Planck's law from those at 
ESPaDOnS wavelengths, which resembles the case of another TTS recently studied with SPIRou \citep[namely V410~Tau,][]{Finociety21}.  The magnetic map we derive in this case (not shown) is similar to that obtained when assuming 
$\delta\simeq0.4$ (although with a $\simeq$20\% weaker dipole component), and so is the map of accretion-induced excess emission in the NC of the \caii\ IRT lines.  The brightness map at ESPaDOnS wavelength is also similar to that shown in 
Fig.~\ref{fig:mapes} (bottom left plot), whereas that at SPIRou wavelength is featureless (by design).  The predicted relative photometric variations corresponding to both, featuring peak-to-peak amplitudes of only 6\% and 0.5\% respectively, 
would again be hardly detectable in the observed light curves of an actively accreting cTTS like CI~Tau.  

One way to distinguish between this second option (where most of the distortions in the LSD profiles of photospheric lines at SPIRou wavelengths are attributed 
to magnetic fields, as in Sec.~\ref{sec:zd1}) and the first one (with $\delta\simeq0.4$, and where brightness spots also contribute to the LSD profile distortions) is to look at LSD profiles of CO molecular lines, in particular those of the 
2.3~\mic\ CO bandhead, that are not sensitive to magnetic fields.  Although the SNR in the LSD profiles associated with CO lines only (not shown) is about twice lower than that when all atomic lines are used, one can nonetheless distinguish profile 
distortions and variability that probe the presence of surface temperature features. For this reason, we conclude that our first option, corresponding to $\delta=0.4$ and to the maps shown in Fig.~\ref{fig:mapes}, is the most likely.  This implies in particular that dark spots do contribute to LSD Stokes $I$ profile distortions, though only marginally in the SPIRou domain, hence why they do not show up in the brightness maps derived from SPIRou data only.  We come back on this point in Sec.~\ref{sec:rvs}.  

We also confirm our preliminary conclusion that the large-scale field of CI~Tau significantly changed since our first study \citep{Donati20b}, and can further claim that the brightness distribution drastically evolved between the two 
epochs, from a dark region at photospheric level that is strongly off-centred with respect to the pole in early 2017 \citep[generating a 4--5~\kms\ amplitude RV signature of the LSD Stokes $I$ profiles, e.g.,][]{Donati20b} to a much more 
polar-centred spot configuration (inducing 10$\times$ smaller peak-to-peak RV variations of 0.5~\kms, see Fig.~\ref{fig:esp} and Sec.~\ref{sec:rvs}).  The same conclusion applies to the reconstructed maps of accretion-induced emission that 
is also much more centred on the pole during our 2020 observing season.  Besides, our results unambiguously confirm that the RV signature with a 9-d period that CI~Tau exhibited in late 2017 can be safely attributed to the activity of the star 
itself rather than to a close-in massive planet \citep{JohnsKrull16, Biddle18, Biddle21}, as further discussed in Sec.~\ref{sec:rvs}.

\section{Veiling and mass accretion rate of CI~Tau}
\label{sec:vei}

Veiling in the spectrum of CI~Tau and other cTTSs observed with SPIRou was recently discussed at length in \citet{Sousa23}.  Their conclusions from the same CI~Tau spectra in seasons 2019 and 2020 are that average veiling in the $Y$, $J$ and $H$ 
bands is of order 0.4--0.5, whereas average veiling in the $K$ band is significantly larger, of order 0.9, with only limited temporal variations.  They find in particular that the nIR veiling reflects dust heating in the inner disc, presumably 
boosted by accretion, hence why veiling is maximum in the K band over the SPIRou domain.  The veiling we derive at optical wavelengths from our ESPaDOnS spectra in season 2020 is in the range 0.8--1.4 (see Fig.~\ref{fig:esp} middle left plot), 
{\emr showing that nIR and optical veiling are different in nature and do not trace the same physical effects and spatial regions in the magnetosphere of CI~Tau.  We find that the optical veiling of CI~Tau, tracing accretion at the surface of the 
star, does not peak when the accretion funnel crosses the line of sight (at phase 0.9, see Sec.~\ref{sec:zd2});  although surprising, this situation can happen when the accretion region is close to the pole \citep[e.g., on DN~Tau in 2012,][]{Donati13}, 
as a likely result of the high level of intrinsic variability inherent to accretion processes.}

Regarding the mass accretion rate \Mdot\ of CI~Tau, \citet{Sousa23} conclude, from the equivalent widths of the \pab\ and \brg\ lines, that $\log\Mdot=-7.5\pm0.2$~\mspy, consistent with the measurement from \hal\ 
and \hbe\ lines in our previous ESPaDOnS campaign \citep[$\log\Mdot=-7.6\pm0.3$~\mspy,][]{Donati20b}.  For optical emission lines that show both a NC and a broad component (BC), namely the \hei\ $D_3$ and the \caii\ IRT lines, the NC, 
tracing the material reaching the stellar surface at the footpoints of accretion funnels, has a much weaker equivalent width \citep[by typically an order of magnitude,][]{Donati20b} than the BC, which is thought to trace material 
further out in the magnetosphere, in the inner disc {\emr or in the stellar wind.  In the particular case of CI~Tau where the BC of the \hei\ $D_3$ line is significantly blue shifted with respect to, and uncorrelated with, the NC \citep{Donati20b}, 
it further suggests that the BC actually traces a hot wind coming from the star itself and / or the inner disc, rather than in-falling material within the funnel flows \citep[][]{Beristain01}.  
In such a case, one may argue that only the NC of these lines, i.e., the one associated with funnel flows, should be taken into account to derive an estimate of \Mdot, and doing so yields an average value of $\log\Mdot=-8.5\pm0.2$~\mspy\ 
for both our 2020 ESPaDOnS data and our older ones from early 2017 \citep[using the same method as][]{Donati20b}. } 
Admittedly, empirical relations between emission line fluxes and accretion luminosities were derived from the whole lines and not from subcomponents, like the NC or BC of the \hei\ $D_3$ and the \caii\ IRT lines;  {\emr however, these relations 
may also overestimate \Mdot\ in cases like CI~Tau where the BC of both lines, and thus the Ba, \pab\ and \brg\ lines as well, are potentially contaminated by a hot wind component from the star and / or the inner disc.   
It may for instance suggest that most of the inner disc material of CI~Tau, potentially contributing to the BC of the \hei\ $D_3$ and the \caii\ IRT lines and to the Ba, \pab\ and \brg\ lines, is either ejected 
through magnetically-driven disc winds \citep[e.g.,][]{Shu94,Nisini04} or magnetospheric ejection events \citep[as in, e.g.,][]{Romanova04, Ustyugova06, Zanni13, Pantolmos20}, with only a small fraction 
reaching the surface of the host star and thereby showing up in the NC of the \hei\ $D_3$ and the \caii\ IRT lines.  The large mass loss rate recently reported from outflows and jets in the inner disc of CI~Tau 
\citep[$\log\Mdot=-7.0$~\mspy,][]{Flores-Rivera23}, consistent with the mass accretion rate derived from the \pab\ and \brg\ lines, comes as further support for our hypothesis.  
This picture, qualitatively consistent with the theoretical model of \citet{Lesur21}, is however still speculative at this stage.  
In the following we simply consider that the two cases mentioned above correspond to an upper and a lower limit on the mass accretion rate at the surface of CI~Tau.  } 

From the simulation results of \citet{Bessolaz08} regarding the radius of the magnetospheric gap \rmag\ that the large-scale magnetic field of CI~Tau is able to carve at the centre of the inner accretion disc at the epoch of our observations, 
and using our estimate of the dipole component of the large-scale field of CI~Tau as determined from our 2020 SPIRou and ESPaDOnS data, we find that $\rmag/\rcor\simeq0.35$ for $\log\Mdot\simeq-7.5$~\mspy\ and  $\rmag/\rcor\simeq0.67$ for 
$\log\Mdot\simeq-8.5$~\mspy, {\emr with typical uncertainties on \rmag/\rcor\ of order 10\%.  Very similar results are obtained when using expressions of the magnetospheric radius from other studies \citep[e.g.,][]{Blinova16,Blinova19,Pantolmos20}.  
In the first case, this ratio is clearly too small for the large-scale magnetic field to trigger the observed stable poleward accretion, and succeed in spinning down CI~Tau to its slow rotation rate \citep[according to simulations, 
e.g.,][]{Zanni13, Blinova16, Pantolmos20};  in the second case however, the large-scale field is just strong enough, yielding a better agreement with observations.  Besides, recent simulations suggest that tilts in the large-scale field of 
the host star, like those we report for CI~Tau, are able to induce warps of the inner accretion disc, but not a significant change in the magnetospheric size \citep[e.g.,][]{Romanova21}.  Reconciling our upper limit on the mass accretion 
rate with simulation predictions would require CI~Tau to host a magnetic field with a dipole component as strong as 3.5~kG, which is clearly inconsistent with existing observations.  Another option is that simulations are grossly off, 
e.g., lacking one critical physical ingredient;  however, it seems unlikely given their relative success at reproducing a variety of observational cases \citep[e.g.,][]{Blinova19}.  } 

\begin{figure*}
\hbox{\includegraphics[scale=0.26,bb=0 0 960 1100]{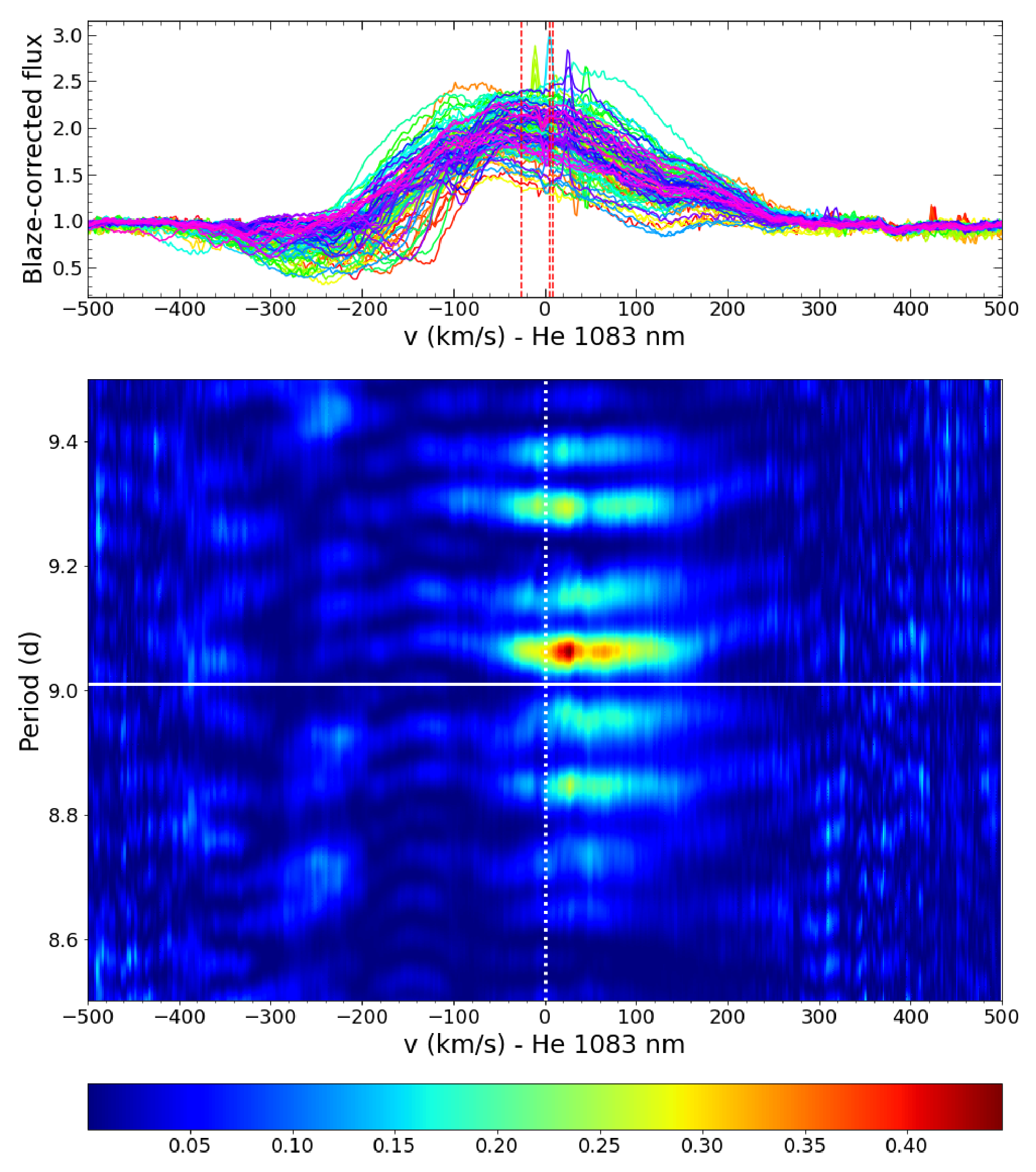}\hspace{2mm}\includegraphics[scale=0.26,bb=0 0 960 1100]{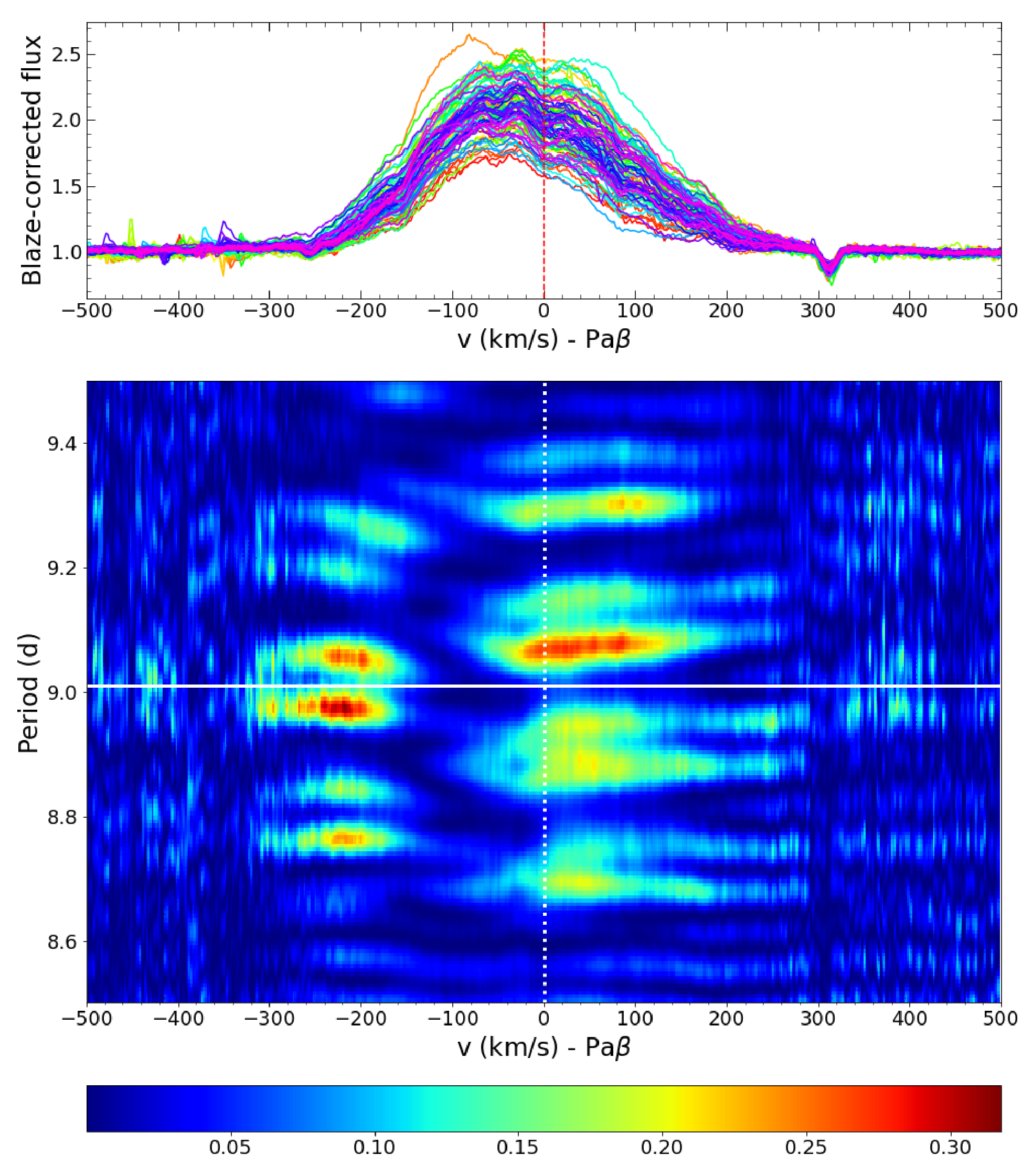}} 
\caption[]{Spectra of CI~Tau in the region of the \hei\ triplet (top left) and \pab\ line (top right), in the stellar rest frame, and the corresponding 2D periodograms (bottom).  
In the two upper panels, the curves show the superposition of all individual spectra, whereas the red vertical dotted lines depict the location of the \hei\ triplet and of the \pab\ line. 
The horizontal white line in the bottom panels indicate the rotation period measured from \Bl\ data.  } 
\label{fig:hpl}
\end{figure*}

Looking at the temporal variability of the 1083~nm \hei, \pab\ and \brg\ lines of CI~Tau over the full set of SPIRou observations yields further information about how accretion proceeds within the magnetosphere of CI~Tau.  The 
1083~nm \hei\ line features a more or less centred emission component with a full-width-at-half-maximum (FWHM) of $\simeq$250~\kms, along with a blue-shifted absorption component extending down to --350~\kms\ (see 
Fig.~\ref{fig:hpl} top left panel).  An additional red-shifted pseudo-absorption component (always staying above a relative flux level of 1.0) also shows up at velocities 0--120~\kms.  All three components are found to vary 
with time.  The corresponding 2D periodogram (see Fig.~\ref{fig:hpl} bottom left panel) features a clear peak at a period of about 9.05~d, i.e., slightly larger than that found from \Bl\ data, in conjunction with the red-shifted absorption 
component.  This red-shifted component most likely traces the outer part of the accretion funnel linking the inner disc to the visible polar regions at the stellar surface of CI~Tau, as it periodically crosses the line of sight.  
Since no clear 9-d periodicity shows up in association with the blue-shifted absorption component of the \hei\ line, which would argue in favour of a stellar wind, we conclude that it presumably probes a wind from the inner regions of the 
accretion disc \citep[despite the overall shape of the \hei\ line may suggest a stellar wind, e.g.,][]{Kwan07}.  In fact, the 2D \hei\ periodogram only shows power in the blue wing for periods larger than $\simeq$17~d (see 
Fig.~\ref{fig:perf} left panel), suggesting a structured disc wind from radii 0.13~au and beyond.  

The \pab\ line consists mainly of a central emission of FWHM $\simeq$250~\kms\ (as for the central emission of the 1083~nm \hei\ line) {\emr with a pseudo-absorption feature centred at velocities of about 0--120~\kms\ 
(see Fig.~\ref{fig:hpl} top right panel).  The 2D periodogram shows that it apparently varies with the same period as the red-shifted absorption component of the 1083~nm \hei\ line 
(see Fig.~\ref{fig:hpl} bottom right panel) and thereby likely traces as well the accretion funnel linking the inner disc to the star as it crosses the line of sight.  Another weaker blue-shifted component at velocities of about --200~\kms} 
also varies with a period of about 8.95~d, i.e., slightly shorter than that found from the \Bl\ data, and possibly with a period of 9.05~d as well (although some of the corresponding power, but not all, may reflect aliasing).  As for \hei, 
the 2D periodogram shows most power in the blue wing of \pab\ for periods larger than $\simeq$17~d (see Fig.~\ref{fig:perf} right panel), suggesting again a structured disc wind from radii 0.13~au and beyond.  
Finally, we find that \brg\ (not shown) exhibits an FWHM and temporal behaviour similar to those of \pab\ but with more noise, making it harder to diagnose the main components of the line profile and their corresponding periodicities.  

Altogether, these lines indicate that the outer portions of accretion funnels (generating the red-shifted absorption component) are modulated on a timescale that is slightly longer than the rotation period measured from \Bl\ data, 
possibly because they are anchored in a region of the inner disc that rotates a bit more slowly than the star and thereby forces accretion funnels to trail stellar rotation;  however, this would contradict our previous 
finding that \rmag\ is unlikely to be larger than \rcor\ given the derived magnetic field and mass accretion rates.  Another option is that CI~Tau exhibits surface differential 
rotation, with the polar regions, where accretion funnels are anchored, rotating more slowly than the average rotation rate at the surface of the star (as measured from \Bl).  That the weak blue-shifted pseudo-absorption component 
of \pab, presumably tracing a stellar wind, is also modulated, with two different periods that are slightly shorter and slightly longer than the average stellar rotation period respectively, may further support this second option and 
suggest the presence of a weak accretion-powered stellar wind emerging from both low and high latitudes at the surface of CI~Tau, in a way similar to coronal streamers and coronal holes in the case of the solar wind.  This option would 
imply a surface differential rotation of order 10~\mrpd\ at the surface of CI~Tau, consistent with the low differential rotation latitudinal shear measured in other fully convective TTSs \citep[e.g.,][]{Finociety23} and justifying 
independently our choice of assuming solid body rotation in the imaging section of our study (see Sec.~\ref{sec:zdi}).  

More generally, the 9-d periodicity that the \hei\ and \pab\ lines exhibit on a time scale of 1200~d in the red wing (and the blue wing as well in the case of \pab) demonstrates that magnetospheric accretion on CI~Tau proceeds in 
a relatively stable manner along a more or less stable magnetic topology, rather than in an unstable fashion through chaotic and rapidly evolving accretion tongues crossing field lines towards the stellar equator \citep[e.g.,][]{Blinova16}, as 
no clear period would otherwise emerge from the 2D periodograms.  This further supports our previous conclusion that $\rmag/\rcor$ must be larger than 0.6 in CI~Tau.  

We show in Fig.~\ref{fig:bal} the stacked \hal\ and \hbe\ profiles of CI~Tau collected during our late 2020 ESPaDOnS run, which also show clear pseudo-absorption features in the red and blue wings of the profiles, qualitatively similar to what we see in \pab.

\section{Radial velocity modeling CI~Tau}
\label{sec:rvs}

Finally, we examine RVs of CI~Tau derived from Gaussian fits to the Stokes $I$ LSD profiles of our SPIRou spectra.  We end up with 94 values covering an overall timescale of about 1200~d, both for our main LSD profiles extracted 
from atomic lines throughout the whole SPIRou domain, and for those derived using the CO bandhead mask (see Sec.~\ref{sec:obs}).  We use simple Gaussian fits to LSD profiles to extract RVs in this study, rather than the 
line-by-line approach \citep{Artigau22} shown to behave optimally for stars whose spectra vary modestly with respect to the median, but to be less suited for accreting cTTSs that exhibit large spectral variations with 
time (including veiling, see Sec.~\ref{sec:vei}).  We find that CI~Tau features a median RV of 16.7~\kms\ with an RMS dispersion of 0.3~\kms\ in atomic lines, and a median RV of 17.1~\kms\ with an RMS dispersion of 0.5~\kms\ in CO 
bandhead lines.  The small RV shift between the two sets of lines may reflect that on average, atomic lines are not formed at the same atmospheric depth as the CO bandhead lines, and therefore do not probe the same average convection 
velocities.  In particular, we speculate that the downward flows in the falling lanes of convective cells are relatively brighter for CO lines than for atomic lines (with respect to the upward flows in the rising granules), thereby 
inducing a small red shift between the two sets of lines.  The larger RV dispersion of CO lines largely reflects the lower SNR of the corresponding LSD profiles (with respect to those from atomic lines).  

We analyse these RVs in the same way as in our previous studies \citep[e.g.,][]{Donati23}, looking for the impact of modulated activity on the RVs but also for the potential contribution of a close-in massive planet that may 
be orbiting around CI~Tau \citep[at a different period than that originally proposed by][as we now know that the 9-d period traces stellar rotation]{JohnsKrull16}.  Following Manick et al.\ (submitted) who find evidence 
for a close-in massive planet on a 24-d orbit around CI~Tau, we look whether an RV signal may be present in the data, in particular at orbital periods close to 24~d.  
We achieve this by adjusting our RV measurements using GPR to model the contribution from stellar activity (as in Sec.~\ref{sec:bl}), and 
a sine wave (or a Keplerian model if needed) to describe the reflex motion induced by a potential close-in planet.  We then run a Monte Carlo Markov Chain (MCMC) experiment to derive optimal values, error bars and posterior 
distributions for all model parameters, in the cases where only activity is taken into account and where both activity and a close-in planet are considered.  In both cases, the evolution timescale ($\theta_3$) and the smoothing 
parameter ($\theta_4$) are fixed to typical values of 200~d and 0.5, to limit the flexibility of GPR given that the error bars on both sets of RVs are only a few times smaller than their RMS dispersion.  The results of the MCMC 
runs are detailed in Table~\ref{tab:pla}.  

We find that stellar activity is contributing to the RV curve at a semi-amplitude level of 0.30~\kms\ for atomic lines and 0.20~\kms\ for CO bandhead lines, demonstrating that magnetic effects indeed dominate profile distortions of 
atomic lines, as anticipated in Sec.~\ref{sec:zdi}.  It also demonstrates that, although smaller, brightness variations at the surface of CI~Tau contribute to profile distortions as well, in particular for CO bandhead lines that are 
not sensitive to magnetic fields.  This further supports our conclusion that CI~Tau hosts cool spots at its surface that are affecting photospheric line profiles, mostly at ESPaDOnS wavelengths but also in the SPIRou domain, consistent with 
our previous conclusion (see Sec.~\ref{sec:zd2}).  We note that the period with which the activity jitter is modulated (equal to $8.92\pm0.02$~d in the case of atomic lines, and a similar value with a larger error bar in the case 
of the CO bandhead lines, see Table~\ref{tab:pla}) is slightly but significantly lower than that derived from \Bl\ data.  As in Sec.~\ref{sec:vei}, we speculate that this difference reflects latitudinal differential rotation at the 
surface of CI~Tau, with regions contributing most to the activity jitter being those located close to the equator, i.e., at a lower latitude than those shaping the \Bl\ curve.  
Incidently, we find that the FWHMs of the LSD profiles of atomic (varying from 21.4 to 25.0~\kms) are also rotationally modulated (the GPR fit yielding an RMS of 0.5~\kms), with a clear period of $8.93\pm0.04$~d, consistent with that 
derived from the associated RVs, and a semi-amplitude of $0.7\pm0.2$~\kms.  Besides, we find that these RVs correlate reasonably well with the first derivative of the FWHMs given the significant level of noise on both quantities 
(correlation coefficient $R\simeq0.5$), further demonstrating that magnetic activity is indeed causing the 9-d RV modulation.  

\begin{figure*}
\centerline{\includegraphics[scale=0.6,angle=-90]{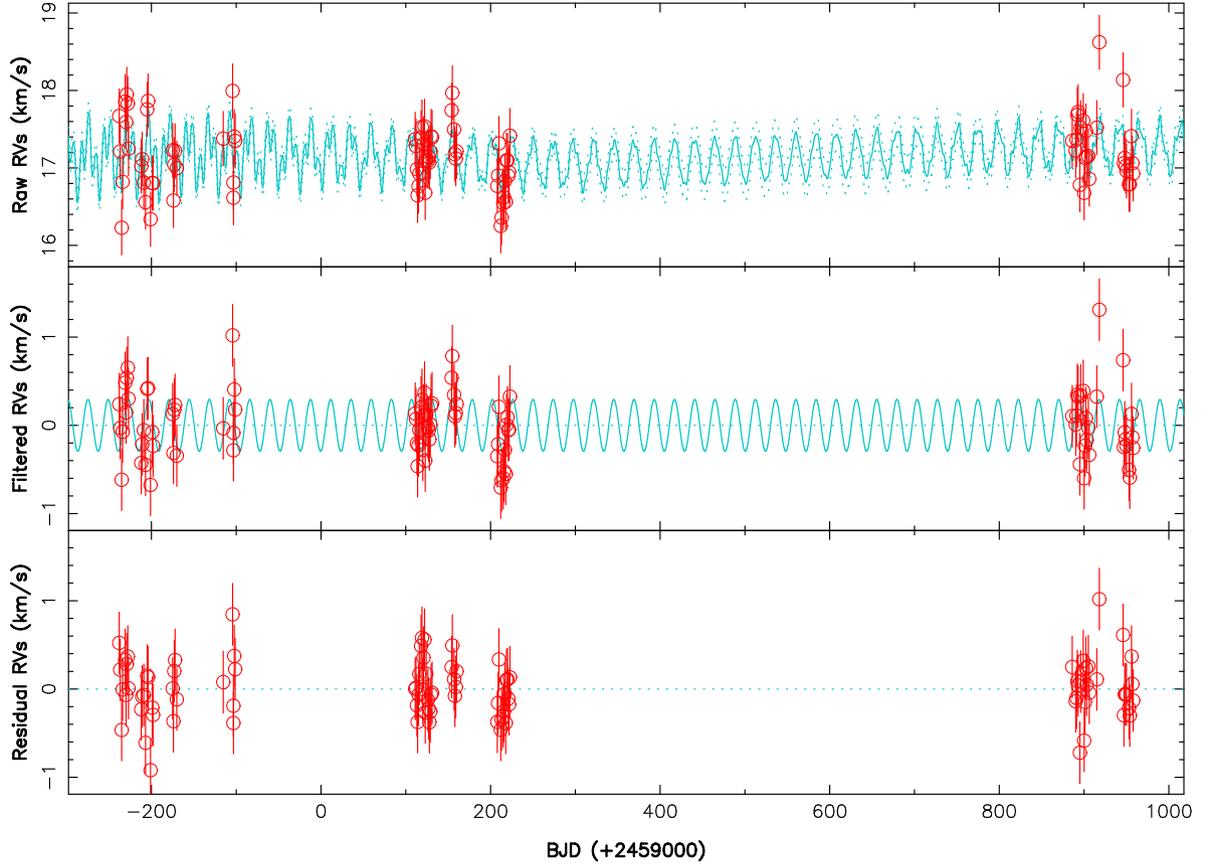}} 
\caption[]{Raw (top), filtered (middle) and residual (bottom) RVs of CI~Tau from CO bandhead lines (red dots) over our SPIRou observing campaign of 1200~d.  The top plot shows the MCMC fit to the RV data, including a QP GPR modeling 
of the activity and the RV signatures of a putative close-in planets of orbital period $23.86\pm0.04$~d (cyan), whereas the middle plot shows the planet RV signature alone once activity is filtered out.  The RMS of the RV residuals is 0.32~\kms. } 
\label{fig:rvr}
\end{figure*}

\begin{figure}
\includegraphics[scale=0.48,angle=-90]{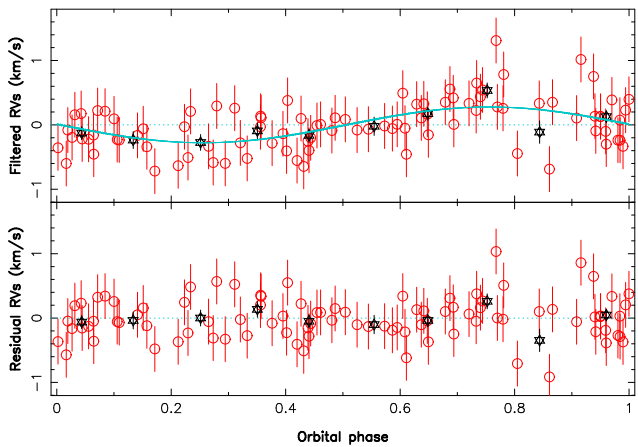}
\caption[]{Filtered (top plot) and residual (bottom plot) RVs for the 23.86-d RV signal detected in CO bandhead lines of CI~Tau.  
The red dots are the individual RV points with the respective error bars, whereas the black stars are average RVs over 0.1 phase bins.
As in Fig.~\ref{fig:rvr}, the dispersion of RV residuals is 0.32~\kms. }
\label{fig:rvf}
\end{figure}

We also examined the possible presence of a close-in massive planet through the reflex motion and the RV signature it may induce on CI~Tau itself, in particular at the orbital period at which a tentative signal has recently been reported 
(i.e., 24~d, Manick et al., submitted) after being initially spotted in photometric data but finally labelled as a likely alias \citep[][]{Biddle18}.  The marginal logarithmic likelihood $\log \mathcal{L}_M$ of a given solution 
is computed using the approach of \citet{Chib01} as described in \citet{Haywood14}, and the significance of the RV signature of a putative planet is estimated from the difference in $\log \mathcal{L}_M$, i.e., the logarithmic Bayes 
Factor $\log$~BF, with respect to our reference model with no planet.  We find that atomic lines show no more than a hint of a signal (at a 2$\sigma$ level) at this period, with an amplitude of only $K_b=0.06^{+0.04}_{-0.03}$~\kms\ 
(corresponding to a minimum planet mass of $M_b \sin i = 0.79^{+0.52}_{-0.39}$~\mjup\ if the RV signal is caused by a planet) and a $\log$~BF of only 1.9, i.e., clearly too small for this signal to be even qualified as tentative.  
Carrying out the same experiment on the RVs from the CO bandhead lines, we obtain a different result, this time with a clear 5.6$\sigma$ signal of semi-amplitude $K_b = 0.28^{+0.06}_{-0.05}$~\kms\ (corresponding to a minimum planet mass 
of $M_b \sin i = 3.70^{+0.80}_{-0.66}$~\mjup) at a well-determined orbital period of $P_b=23.86\pm0.04$~d, and associated with a $\log$~BF of 13.1 apparently indicating a firm detection.  The resulting fit to the raw RVs is shown in 
Fig.~\ref{fig:rvr} (top panel) along with the candidate planet signal in the filtered RVs (middle) and the residual RVs (bottom, RMS of 0.32~\kms), whereas the RV curve phase-folded on the recovered orbital period is displayed in 
Fig.~\ref{fig:rvf}.  Assuming a Keplerian rather than a circular orbit to model the planet signature yields an eccentricity consistent with 0 (with an error bar of about 0.25) and does not improve $\log$~BF in a significant way 
($\Delta \log$~BF~$\simeq0.1$), justifying a posteriori our hypothesis of a circular orbit.  {\emr We point out that, although similar at first glance, the characteristics of the detected RV signal inferred from our study significantly 
differ from those of Manick et al.\ (submitted), whose results we were not able to reproduce, even from their own data. }  
We finally note that no clear signal shows up in the 2D periodograms of \hei, \pab\ and \brg\ lines at the period of the detected RV signature, apart from a weak one in the blue wing of \hei\ (see Fig.~\ref{fig:perf} for the \hei\ and \pab\ lines).   

\begin{table*}
\caption[]{MCMC results for the 4 studied cases (no planet and planet b, using RVs from atomic lines and from CO bandhead lines).  For each case, we list the recovered GP and planet parameters with their error bars, as well 
as the priors used whenever relevant.  The last 4 rows give the \chisqr\ and the RMS of the best fit to our RV data, as well as the associated marginal logarithmic likelihood $\log \mathcal{L}_M$ and marginal logarithmic likelihood 
variation $\Delta \log \mathcal{L}_M$ with respect to the corresponding model without planet. }
\begin{tabular}{ccccccccc}
\hline
                   && \multicolumn{2}{c}{Atomic lines}                   && \multicolumn{2}{c}{CO bandhead}        &&         \\
Parameter          && No planet              & b                         && No planet              & b             &&   Prior \\
\hline
$\theta_1$ (\kms)  && $0.29^{+0.09}_{-0.07}$ & $0.29^{+0.08}_{-0.06}$    && $0.21^{+0.09}_{-0.06}$ & $0.20^{+0.09}_{-0.06}$   && mod Jeffreys ($\sigma_{\rm RV}$) \\
$\theta_2$ (d)     && $8.92\pm0.02$          & $8.92\pm0.02$             && $8.91\pm0.12$          & $8.77\pm0.17$            && Gaussian (9.0, 0.5) \\
$\theta_3$ (d)     && 200                    & 200                       && 200                    & 200                      &&  \\
$\theta_4$         && 0.5                    & 0.5                       && 0.5                    & 0.5                      &&  \\
$\theta_5$ (\kms)  && $0.12\pm0.05$          & $0.11\pm0.05$             && $0.18\pm0.09$          & $0.11\pm0.08$            && mod Jeffreys ($\sigma_{\rm RV}$) \\
\hline
$K_b$ (\kms)       &&                        & $0.06^{+0.04}_{-0.03}$    &&                        & $0.28^{+0.06}_{-0.05}$   && mod Jeffreys ($\sigma_{\rm RV}$) \\
$P_b$ (d)          &&                        & $23.79\pm0.15$            &&                        & $23.86\pm0.04$           && Gaussian (23.80, 1.0) \\
BJD$_b$ (2459000+) &&                        & $398.2\pm2.3$             &&                        & $398.7\pm0.7$            && Gaussian (398, 5) \\
$M_b \sin i$ (\mjup) &&                      & $0.79^{+0.52}_{-0.39}$    &&                        & $3.70^{+0.80}_{-0.66}$   && derived from $K_b$, $P_b$ and \mstar \\
\hline
\chisqr            && 1.14                   & 1.02                      && 1.12                   & 0.86                     &&  \\
RMS (\kms)         && 0.19                   & 0.18                      && 0.37                   & 0.32                     &&  \\
$\log \mathcal{L}_M$ && --6.7                & --4.8                     && --51.7                 & --38.6                   &&  \\
$\log {\rm BF} = \Delta \log \mathcal{L}_M$ && 0.0 & 1.9                 && 0.0                    & 13.1                     &&  \\
\hline
\end{tabular}
\label{tab:pla}
\end{table*}

At this point, it is not clear why the detected signal, if induced by an orbiting planet, would be detected in the RVs of the CO bandhead lines but not in those of atomic lines, despite the latter being more precise and less dispersed 
around the GPR fit (0.18~\kms\ RMS) than the former (0.32~\kms\ RMS).  If probing the presence of a close-in planet, the RV signal should indeed show up in the same way in both sets of lines, and not just in one of them.  Admittedly, 
CO bandhead lines are less affected by activity, in particular magnetic fields;  however, the RV precision that they yield is twice lower than that from atomic lines, implying that a putative planet signal that shows up in CO lines should 
also be detected with a similar semi-amplitude in atomic lines.  We speculate that the detected RV signal rather probes the presence of a non-axisymmetric density structure in the accretion disc, located at an average distance of $\simeq$0.16~au 
where the Keplerian period is consistent with that of the detected RV signal ($23.86\pm0.04$~d), and distorting the Stokes $I$ LSD profiles of atomic and CO bandhead lines of CI~Tau in a different way as a result of, e.g., the disc structure 
being much cooler than the photosphere of CI~Tau.  Although the Keplerian velocity at a distance of 0.16~au (i.e., 57.4~\kms) is in principle 2.5$\times$ and 3.8$\times$ larger than the FWHM of the LSD Stokes $I$ profiles of CI~Tau in 
atomic and CO bandhead lines respectively, some of the disc material may rotate at a strongly sub-Keplerian velocity as a result of, e.g., magnetic fields \citep[][]{Donati05, Shu07}, thereby contributing to the disc profile with a component 
whose FWHM is closer to that of the stellar lines.  As this non-axisymmetric disc structure orbits around the star, its spectral contribution, adding up to the stellar spectrum, is expected to modulate RVs of LSD profiles in a way that 
depends on the selected lines, qualitatively similar to what we observe.  The nature of this axisymmetric structure is unclear, but could reflect the presence of strong magnetic fields in the disc, as in the case of 
FU~Ori \citep[][]{Donati05};  it may also relate to the presence of a vortex \citep{Varga21}, and / or a nascent planet, with a mass low enough not to be detected in atomic lines but large enough to induce a non-axisymmetric density structure in the accretion disc.  

We note that the beat periods between the 23.86-d RV signal and the 9.01-d rotation period of CI~Tau, equal to 6.5 and 14.5~d, fall close to the secondary peaks in the periodogram of the 2017 K2 light curve of CI~Tau as reported 
by \citet[][located at 6.6 and 14.3~d]{Biddle18}.  It may suggest that the putative disc structure at 0.16~au that we invoke to explain the RV signal in CO bandhead lines may also participate to star-disc interactions, e.g., generating 
enhanced accretion each time it falls in conjunction with the accretion funnels that rotate with the star {\emr \citep[e.g., as in the pulsed accretion model of][]{Teyssandier20}}.

\section{Summary and discussion}
\label{sec:dis}

In our paper we report new observations of the cTTS CI~Tau collected with the SPIRou nIR spectropolarimeter / velocimeter at CFHT in the framework of the SLS and SPICE large programmes, complemented with PI observations recorded with the 
ESPaDOnS optical spectropolarimeter.  A total of 94 and 13 validated Stokes $I$ and $V$ spectra of CI~Tau were collected with SPIRou and ESPaDOnS, over a period of about 1200~d spanning three observing seasons for SPIRou (2019, 2020 and 2022), 
and over 16~d in late 2020 for ESPaDOnS.  LSD profiles were computed for all spectra, yielding \Bl\ values that are unambiguously modulated by the rotation of the star, equal to $9.010\pm0.023$~d.  We can thus safely exclude the alternate 
rotation period (of 6.6~d) that was initially proposed as the rotation period of CI~Tau in several studies \citep{JohnsKrull16,Biddle18,Biddle21}, suggesting that this alternate period may relate to some other phenomenon of the complex 
star-disc interactions taking place in CI~Tau.  

We observe that \Bl\ curves differ in both wavelength domains, with nIR photospheric lines probing longitudinal fields varying from --170 to 120~G over our three observing seasons (including 2020), whereas optical photospheric lines 
yield \Bl\ values only fluctuating from --5 to 130~G in late 2020.  It indicates the presence of brightness features at the surface of CI~Tau with a contrast that depends on wavelength, thereby affecting differently optical and nIR spectra.  
Modeling the Stokes $I$ and $V$ profiles of SPIRou spectra with ZDI allows us to reconstruct the distribution of brightness and magnetic features at the surface of CI~Tau, indicating that CI~Tau hosts a mainly poloidal and axisymmetric 
large-scale field whose average strength ranges from 0.8 to 0.9~kG, and that essentially consists of a slightly tilted dipole with a polar strength of 0.8~kG.  Adding in the ESPaDOnS data for the ZDI modeling of our 2020 observations provides 
stronger constraints on the field topology, the associated brightness map and the distribution of the accretion-induced excess emission in \caii\ IRT lines.  It tells that accretion mainly proceeds towards the pole at the surface of CI~Tau, 
with the footpoints of accretion funnels showing up as bright features at chromospheric level and dark features at photospheric level.  The updated magnetic topology is similar to that derived from SPIRou data only, except for a stronger 
dipole component (to compensate for the relative darkness of polar regions) of strength 1.1~kG.  We also report a clear change of both the magnetic topology and brightness map of CI~Tau with respect to our previous ESPaDOnS observations in 
early 2017 \citep{Donati20b}, with the dark features at the footpoints of accretion funnels being now much less off-centred with respect to the pole and thereby generating much smaller profile distortions and RV variations (in both optical 
and nIR domains).  

Both the 1083~nm \hei\ and \pab\ lines show a clear modulation at a 9-d period in the red wing of their profile, tracing the base of the accretion funnel close to the inner disc crossing the line of sight as the star rotates.  The blue wing 
of \pab\ is also modulated by rotation, possibly probing a weak accretion-powered stellar wind escaping from both low and high latitude regions at the surface of the star.  These modulation periods, slightly different from that of \Bl\ data, suggest that 
surface differential rotation is weak at the surface of CI~Tau, consistent with previous results on fully convective TTSs \citep[e.g.,][]{Finociety23}.  Altogether, it indicates that accretion is stable enough at the surface of CI~Tau to generate a clear 
signal in the 2D periodograms of both lines, since unstable accretion as described by, e.g., \citet[][]{Blinova16}, with disc material penetrating the magnetosphere through chaotic equatorial accretion tongues, is unlikely to produce a coherent signal on a 
timescale of 1200~d.  The mass accretion rate inferred from the \pab\ and \brg\ lines of the same SPIRou spectra by \citet{Sousa23} {\emr is equal to $\log \Mdot=-7.5\pm0.2$~\mspy, similar to that derived from Ba lines in optical spectra 
\citep{Donati20b}.  In the particular case of CI~Tau however, the BC of the \hei\ $D_3$ line is significantly blue shifted, suggesting that it probes a hot wind from the star and / or the inner accretion disc rather 
than in-falling material in funnel flows \citep{Beristain01}.  Considering the NC of both features only yields $\log \Mdot=-8.5\pm0.2$~\mspy\ for CI~Tau in 2020, which we consider as a lower limit to the mass accretion rate, as opposed to the 
upper limit derived from Ba, \pab\ and \brg\ lines, or from the combined NC and BC of the \hei\ $D_3$ and \caii\ IRT lines.  
We find that this upper limit implies a magnetospheric size (relative to the corotation radius) of about $\rmag/\rcor\simeq0.35$ given the strength we measure for the large-scale dipole field of CI~Tau, whereas the lower limit yields 
$\rmag/\rcor\simeq0.67$.  The former estimate appears inconsistent with requirements for stable accretion as derived from numerical simulations \citep[e.g.,][]{Blinova16,Pantolmos20}, whereas the latter is in better agreement.  Our 
results may thus imply that only a small fraction of the disc material traced with Ba, \pab\ and \brg\ lines reaches the magnetosphere of CI~Tau \citep[as in, e.g.,][]{Lesur21} and makes it to the surface of the star in a stable accretion 
pattern, where it shows up as the NC of the \hei\ $D_3$ and \caii\ IRT lines, whereas the BC of both lines as well as most of material traced with Ba, \pab\ and \brg\ lines would probe a wind from the inner disc and / or the star.  The large 
mass loss rate recently reported from outflows and jets in the inner disc of CI~Tau \citep{Flores-Rivera23}, consistent with the mass accretion rate derived from the \pab\ and \brg\ lines, comes as further support for this hypothesis.  
We however stress that this picture is still speculative and requires confirmation from further observations. } 

We also studied RVs from SPIRou spectra of CI~Tau, analyzing both LSD profiles of atomic lines and those of CO bandhead lines, and using GPR to model the activity jitter.  We find that both sets of RVs are modulated with a period slightly 
smaller than that found from the \Bl\ curve, again suggesting that weak surface differential rotation is present at the surface of the star.  Besides, the semi-amplitude of the modulation, respectively equal to 0.3 and 0.2~\kms\ for atomic 
and CO lines, is inconsistent with the 1~\kms\ semi-amplitude reported for the putative close-in planet \citep{JohnsKrull16}, further demonstrating that the 9-d RV signature is indeed of stellar rather than planetary origin.  Looking 
for RV signatures potentially adding up on top of the activity jitter at a different period, and in particular at the 24~d period recently reported to show up in CI~Tau photometric light curves and SPIRou and ESPaDOnS data 
(Manick et al, submitted), we find 
that atomic lines exhibit no more than a weak, barely significant, 2$\sigma$ signature of semi-amplitude $K_b=0.06^{+0.04}_{-0.03}$~\kms, whereas CO lines show a clear 5.6$\sigma$ signal of semi-amplitude $K_b = 0.28^{+0.06}_{-0.05}$~\kms, 
the latter at a well-defined period of $23.86\pm0.04$~d.  Atomic lines being more affected by activity than CO lines (hence the 1.5$\times$ larger semi-amplitude associated with the activity jitter), one can expect the former to be less sensitive to 
the planet signal than the latter;  however, RVs from atomic lines are about twice more precise than those from CO lines, which should compensate for the lower sensitivity.  For this reason, we speculate that the observed 
23.86-d RV signal rather probes the presence of a non-axisymmetric structure in the accretion disc, possibly linked with strong magnetic fields, a vortex, and / or of a lower-mass planet, at an approximate distance of 0.16~au from the 
centre of the host star.  
This is at least qualitatively consistent with the latest interferometric measurements with GRAVITY \citep{Gravity23}, suggesting the presence of a gap ranging from 0.1 to 0.2~au in the inner disc of CI~Tau, and with the study of CO lines in the M  
band \citep[at 4.7~\mic,][]{Kozdon23}, also arguing for a non-axisymmetric disc structure and a potential gap at a distance of about 0.14~au.   

Our study brings us a step forward in the understanding of the complex magnetospheric accretion processes at work in the star-disc interactions of CI~Tau, and more generally on their role in star / planet formation.  More analyses and observations 
of the same type are nonetheless needed to progress further \citep[e.g.,][]{Bouvier23}, and determine more accurately how much of the inner disc material is being transferred to the magnetosphere or expelled outwards through disc winds.  Similarly, we need to 
understand in a more quantitative way the physical nature of the non-axisymmetric disc structure inducing the observed RV signal detected in CO lines but not in atomic lines of CI~Tau.  Looking in more details at the profiles of CO lines and how they 
vary with time, as in \citet{Flagg19} but for the RV signal we detect in the SPIRou data, should tell us more about what this enigmatic disc structure consists of;  this is already planned for a forthcoming companion paper.   
Our study also 
demonstrates the extreme value of collecting (quasi) simultaneous optical and nIR spectropolarimetric data of cTTSs.  This should be feasible soon with the VISION project at CFHT, that will allow both SPIRou and ESPaDOnS to observe the 
same object with the two instruments, maintaining accurate polarimetric information in both channels. Similarly, coordinated spectropolarimetric and interferometric observations, like those recently carried out with GRAVITY \citep{Gravity23} but 
with a more extensive monitoring of the rotation cycle, should also bring key complementary information about how magnetospheric accretion proceeds in CTTSs like CI~Tau.

\section*{Acknowledgements}
This project received funds from the European Research Council (ERC) under the H2020 research \& innovation program (grant agreement
\#740651 NewWorlds) and under the Horizon Europe research \& innovation program (\#101053020  Dust2Planets), the Agence Nationale pour la Recherche (ANR, project ANR-18-CE31-0019 SPlaSH) and
the Investissements d'Avenir program (project ANR-15-IDEX-02).   SHPA acknowledge funding from FAPEMIG, CNPq and CAPES.  

Our study is based on data obtained at the CFHT, operated by the CNRC (Canada), INSU/CNRS (France) and the University of Hawaii.
The authors wish to recognise and acknowledge the very significant cultural role and reverence that the summit of Maunakea has always had
within the indigenous Hawaiian community.  We are most fortunate to have the opportunity to conduct observations from this mountain.
This work also benefited from the SIMBAD CDS database at URL {\tt http://simbad.u-strasbg.fr/simbad} and the ADS system at URL {\tt https://ui.adsabs.harvard.edu}.

\section*{Data availability}  Data used in this paper are publicly available from the Canadian Astronomy Data Center (CADC).   

\bibliography{citau2} 
\bibliographystyle{mnras}

\appendix
\section{Observation log}
\label{sec:appA}

Tables~\ref{tab:lgs} and \ref{tab:lge} give the full log and associated \Bl\ and RV measurements at each observing epoch, from SPIRou and ESPaDOnS spectra respectively.  

\begin{table*} 
\small
\caption[]{Log of our SPIRou observations of CI~Tau in seasons 2019, 2020 and 2022.  All exposures consist of 4 sub-exposures of equal length.
For each visit, we list the barycentric Julian date BJD, the UT date, the rotation cycle c and phase $\phi$ (computed as indicated in Sec.~\ref{sec:obs}), 
the total observing time t$_{\rm exp}$, the peak SNR in the spectrum (in the $H$ band) per 2.3~\kms\ pixel, the noise level in the LSD Stokes $V$ profile, 
the estimated \Bl\ with error bars, and the RV and error bar of atomic lines and of the 2.3~\mic\ CO bandhead lines. } 
\hspace{-6mm}
\begin{tabular}{ccccccccc} 
\hline
BJD        & UT date & c / $\phi$ & t$_{\rm exp}$ & SNR & $\sigma_V$            & \Bl\   &  RV    &  RV CO   \\ 
(2459000+) &         &            &   (s)        & ($H$) & ($10^{-4} I_c$)       & (G)   & (\kms) & (\kms) \\  
\hline
-237.9633975 & 05 Oct 2019 & 0 / 0.004 & 1582.4 & 204 & 3.12 & 22.4$\pm$16.0 & 17.09$\pm$0.18 & 17.67$\pm$0.35\\
-236.8879737 & 06 Oct 2019 & 0 / 0.123 & 1582.4 & 193 & 3.24 & 42.3$\pm$16.1 & 16.33$\pm$0.18 & 17.21$\pm$0.35\\
-235.0846824 & 08 Oct 2019 & 0 / 0.324 & 1582.4 & 178 & 3.68 & 88.9$\pm$19.6 & 16.58$\pm$0.18 & 16.23$\pm$0.35\\
-233.9109268 & 09 Oct 2019 & 0 / 0.454 & 1582.4 & 177 & 3.51 & -33.9$\pm$16.1 & 16.57$\pm$0.18 & 16.82$\pm$0.35\\
-230.9560376 & 12 Oct 2019 & 0 / 0.782 & 1582.4 & 193 & 3.37 & -17.2$\pm$14.9 & 16.84$\pm$0.18 & 17.86$\pm$0.35\\
-229.9311828 & 13 Oct 2019 & 0 / 0.896 & 1582.4 & 189 & 3.36 & -3.3$\pm$16.0 & 17.08$\pm$0.18 & 17.59$\pm$0.35\\
-228.9897459 & 14 Oct 2019 & 1 / 0.000 & 1582.4 & 206 & 3.46 & 14.2$\pm$17.9 & 16.84$\pm$0.18 & 17.95$\pm$0.35\\
-227.8758917 & 15 Oct 2019 & 1 / 0.124 & 1582.4 & 165 & 4.04 & 20.9$\pm$19.6 & 16.66$\pm$0.18 & 17.83$\pm$0.35\\
-227.0116257 & 16 Oct 2019 & 1 / 0.220 & 1582.4 & 204 & 3.37 & 70.5$\pm$16.8 & 16.38$\pm$0.18 & 17.25$\pm$0.35\\
-211.9422929 & 31 Oct 2019 & 2 / 0.892 & 1582.4 & 191 & 3.46 & 13.1$\pm$17.1 & 16.78$\pm$0.18 & 17.02$\pm$0.35\\
-210.9365839 & 01 Nov 2019 & 3 / 0.004 & 1582.4 & 200 & 3.52 & 21.0$\pm$18.0 & 16.62$\pm$0.18 & 17.11$\pm$0.35\\
-209.0012681 & 03 Nov 2019 & 3 / 0.219 & 1582.4 & 179 & 3.57 & 73.2$\pm$16.6 & 16.35$\pm$0.18 & 16.81$\pm$0.35\\
-206.9469354 & 05 Nov 2019 & 3 / 0.447 & 1582.4 & 188 & 3.39 & -48.5$\pm$20.1 & 16.23$\pm$0.18 & 16.56$\pm$0.35\\
-204.9947500 & 07 Nov 2019 & 3 / 0.663 & 1582.4 & 158 & 4.19 & -52.4$\pm$22.4 & 16.98$\pm$0.18 & 17.76$\pm$0.35\\
-203.8616552 & 08 Nov 2019 & 3 / 0.789 & 1582.4 & 134 & 5.48 & -27.3$\pm$27.0 & 16.99$\pm$0.18 & 17.87$\pm$0.35\\
-200.9838626 & 11 Nov 2019 & 4 / 0.108 & 1582.4 & 170 & 4.17 & 51.0$\pm$24.8 & 16.52$\pm$0.18 & 16.34$\pm$0.35\\
-199.0534922 & 13 Nov 2019 & 4 / 0.323 & 1582.4 & 179 & 3.74 & 56.6$\pm$23.2 & 17.06$\pm$0.18 & 16.81$\pm$0.35\\
-198.0391714 & 14 Nov 2019 & 4 / 0.435 & 1582.4 & 196 & 3.39 & -48.9$\pm$18.2 & 16.17$\pm$0.18 & 16.80$\pm$0.35\\
-174.9998742 & 07 Dec 2019 & 6 / 0.992 & 1582.4 & 146 & 4.71 & 57.8$\pm$25.9 & 16.27$\pm$0.18 & 17.23$\pm$0.35\\
-174.0618588 & 08 Dec 2019 & 7 / 0.096 & 1582.4 & 179 & 3.52 & 12.7$\pm$20.1 & 16.27$\pm$0.18 & 16.58$\pm$0.35\\
-173.0536671 & 09 Dec 2019 & 7 / 0.208 & 1582.4 & 176 & 3.63 & 9.6$\pm$20.0 & 16.61$\pm$0.18 & 17.04$\pm$0.35\\
-172.1021738 & 10 Dec 2019 & 7 / 0.314 & 1582.4 & 206 & 3.06 & -1.8$\pm$15.9 & 16.65$\pm$0.18 & 17.22$\pm$0.35\\
-170.0524849 & 12 Dec 2019 & 7 / 0.541 & 1582.4 & 176 & 3.65 & -90.5$\pm$22.1 & 16.33$\pm$0.18 & 17.00$\pm$0.35\\
\hline
-115.2469545 & 05 Feb 2020 & 13 / 0.624 & 1582.4 & 212 & 2.87 & -34.2$\pm$15.8 & 16.92$\pm$0.18 & 17.38$\pm$0.35\\
-104.2449281 & 16 Feb 2020 & 14 / 0.845 & 1582.4 & 171 & 3.37 & -6.4$\pm$20.3 & 17.53$\pm$0.18 & 17.99$\pm$0.35\\
-103.2116761 & 17 Feb 2020 & 14 / 0.960 & 1582.4 & 156 & 3.75 & 49.9$\pm$22.1 & 16.38$\pm$0.18 & 16.81$\pm$0.35\\
-103.1907509 & 17 Feb 2020 & 14 / 0.962 & 1582.4 & 153 & 3.99 & 10.0$\pm$23.6 & 16.43$\pm$0.18 & 16.62$\pm$0.35\\
-102.2435988 & 18 Feb 2020 & 15 / 0.067 & 1582.4 & 188 & 3.06 & 19.2$\pm$18.8 & 16.21$\pm$0.18 & 17.40$\pm$0.35\\
-101.2222892 & 19 Feb 2020 & 15 / 0.181 & 1582.4 & 200 & 2.85 & 12.1$\pm$16.4 & 16.56$\pm$0.18 & 17.35$\pm$0.35\\
\hline
111.0501499 & 18 Sep 2020 & 38 / 0.740 & 1604.7 & 189 & 3.43 & -115.5$\pm$18.4 & 16.66$\pm$0.18 & 17.37$\pm$0.35\\
112.0890261 & 19 Sep 2020 & 38 / 0.856 & 1465.4 & 144 & 4.91 & -44.9$\pm$24.5 & 16.96$\pm$0.18 & 17.28$\pm$0.35\\
113.1107211 & 20 Sep 2020 & 38 / 0.969 & 1604.7 & 227 & 2.69 & -34.1$\pm$12.9 & 16.75$\pm$0.18 & 16.98$\pm$0.35\\
114.0167967 & 21 Sep 2020 & 39 / 0.070 & 1604.7 & 200 & 3.06 & -47.3$\pm$14.7 & 16.38$\pm$0.18 & 16.65$\pm$0.35\\
115.0794741 & 22 Sep 2020 & 39 / 0.188 & 1604.7 & 213 & 2.78 & 12.8$\pm$13.7 & 16.54$\pm$0.18 & 16.77$\pm$0.35\\
116.0763840 & 23 Sep 2020 & 39 / 0.298 & 1604.7 & 211 & 2.91 & 22.8$\pm$13.6 & 16.63$\pm$0.18 & 16.93$\pm$0.35\\
118.0533393 & 25 Sep 2020 & 39 / 0.518 & 1604.7 & 231 & 2.59 & 48.0$\pm$11.8 & 17.00$\pm$0.18 & 17.39$\pm$0.35\\
119.1407575 & 26 Sep 2020 & 39 / 0.638 & 1604.7 & 208 & 2.86 & -50.9$\pm$14.4 & 17.25$\pm$0.18 & 17.53$\pm$0.35\\
120.1241333 & 27 Sep 2020 & 39 / 0.747 & 1604.7 & 125 & 6.17 & -65.8$\pm$30.7 & 16.13$\pm$0.18 & 16.94$\pm$0.35\\
120.9924616 & 28 Sep 2020 & 39 / 0.844 & 1604.7 & 219 & 2.75 & -108.6$\pm$14.4 & 16.88$\pm$0.18 & 17.31$\pm$0.35\\
122.0961681 & 29 Sep 2020 & 39 / 0.966 & 1604.7 & 208 & 3.02 & -101.3$\pm$15.3 & 17.11$\pm$0.18 & 17.54$\pm$0.35\\
122.9866365 & 30 Sep 2020 & 40 / 0.065 & 1604.7 & 238 & 2.62 & -64.4$\pm$14.1 & 16.36$\pm$0.18 & 16.68$\pm$0.35\\
124.0760021 & 01 Oct 2020 & 40 / 0.186 & 1604.7 & 201 & 3.10 & -10.0$\pm$16.1 & 16.62$\pm$0.18 & 17.10$\pm$0.35\\
126.1413499 & 03 Oct 2020 & 40 / 0.415 & 1604.7 & 208 & 2.86 & 73.2$\pm$14.7 & 16.92$\pm$0.18 & 17.09$\pm$0.35\\
126.9898198 & 04 Oct 2020 & 40 / 0.509 & 1604.7 & 218 & 2.77 & 47.5$\pm$13.5 & 16.89$\pm$0.18 & 17.14$\pm$0.35\\
127.9633281 & 05 Oct 2020 & 40 / 0.617 & 1604.7 & 230 & 2.71 & -22.2$\pm$13.8 & 16.89$\pm$0.18 & 17.07$\pm$0.35\\
129.0246679 & 06 Oct 2020 & 40 / 0.735 & 1604.7 & 214 & 3.08 & -74.7$\pm$16.0 & 16.75$\pm$0.18 & 17.21$\pm$0.35\\
129.9552595 & 07 Oct 2020 & 40 / 0.839 & 1604.7 & 237 & 2.64 & -86.4$\pm$13.6 & 16.94$\pm$0.18 & 17.41$\pm$0.35\\
131.0743781 & 08 Oct 2020 & 40 / 0.963 & 1604.7 & 221 & 2.92 & -84.3$\pm$14.5 & 16.80$\pm$0.18 & 17.40$\pm$0.35\\
154.0788176 & 31 Oct 2020 & 43 / 0.516 & 1604.7 & 234 & 2.58 & 36.6$\pm$13.2 & 16.92$\pm$0.18 & 17.75$\pm$0.35\\
154.9587953 & 01 Nov 2020 & 43 / 0.614 & 1604.7 & 244 & 2.47 & -0.8$\pm$12.9 & 16.99$\pm$0.18 & 17.97$\pm$0.35\\
156.9939056 & 03 Nov 2020 & 43 / 0.840 & 1604.7 & 251 & 2.56 & -88.4$\pm$15.6 & 16.86$\pm$0.18 & 17.50$\pm$0.35\\
157.9993166 & 04 Nov 2020 & 43 / 0.951 & 1604.7 & 217 & 2.98 & -129.7$\pm$17.3 & 16.71$\pm$0.18 & 17.18$\pm$0.35\\
159.0002423 & 05 Nov 2020 & 44 / 0.062 & 1604.7 & 240 & 2.64 & -57.0$\pm$14.2 & 16.46$\pm$0.18 & 17.12$\pm$0.35\\
160.0429516 & 06 Nov 2020 & 44 / 0.178 & 1604.7 & 175 & 3.79 & 9.6$\pm$20.5 & 16.57$\pm$0.18 & 17.22$\pm$0.35\\
\hline
\end{tabular} 
\end{table*}

\setcounter{table}{0}
\begin{table*}
\caption[]{continued} 
\hspace{-6mm} 
\begin{tabular}{ccccccccc} 
\hline
BJD        & UT date & c / $\phi$ & t$_{\rm exp}$ & SNR & $\sigma_V$            & \Bl\  &  RV     &  RV CO   \\ 
(2459000+) &         &            &   (s)        & ($H$) & ($10^{-4} I_c$)       & (G)   & (\kms) &  (\kms)  \\  
\hline
207.9508599 & 24 Dec 2020 & 49 / 0.495 & 1604.7 & 230 & 2.67 & 38.7$\pm$14.2 & 17.05$\pm$0.18 & 16.77$\pm$0.35\\
208.9595835 & 25 Dec 2020 & 49 / 0.607 & 1604.7 & 217 & 2.91 & 25.6$\pm$15.2 & 16.60$\pm$0.18 & 16.90$\pm$0.35\\
209.9135399 & 26 Dec 2020 & 49 / 0.713 & 1604.7 & 237 & 2.58 & -79.8$\pm$14.0 & 16.69$\pm$0.18 & 17.32$\pm$0.35\\
211.9861378 & 28 Dec 2020 & 49 / 0.943 & 1604.7 & 186 & 3.55 & -94.0$\pm$20.3 & 16.74$\pm$0.18 & 16.25$\pm$0.35\\
212.9626224 & 29 Dec 2020 & 50 / 0.051 & 1604.7 & 191 & 3.43 & -32.6$\pm$18.7 & 16.44$\pm$0.18 & 16.36$\pm$0.35\\
214.9232126 & 31 Dec 2020 & 50 / 0.269 & 1604.7 & 203 & 3.09 & 43.5$\pm$15.8 & 16.57$\pm$0.18 & 16.55$\pm$0.35\\
215.8398527 & 01 Jan 2021 & 50 / 0.371 & 1604.7 & 184 & 3.59 & 53.6$\pm$18.4 & 16.73$\pm$0.18 & 16.62$\pm$0.35\\
216.8766784 & 02 Jan 2021 & 50 / 0.486 & 1604.7 & 213 & 2.87 & 67.6$\pm$13.9 & 17.03$\pm$0.18 & 16.83$\pm$0.35\\
217.9155611 & 03 Jan 2021 & 50 / 0.601 & 1604.7 & 204 & 2.96 & 3.0$\pm$14.7 & 16.54$\pm$0.18 & 16.56$\pm$0.35\\
218.8995204 & 04 Jan 2021 & 50 / 0.710 & 1604.7 & 208 & 2.98 & -86.3$\pm$14.2 & 16.55$\pm$0.18 & 17.10$\pm$0.35\\
219.9298890 & 05 Jan 2021 & 50 / 0.825 & 1604.7 & 199 & 3.11 & -135.0$\pm$15.7 & 16.72$\pm$0.18 & 17.10$\pm$0.35\\
220.8893133 & 06 Jan 2021 & 50 / 0.931 & 1604.7 & 208 & 2.87 & -108.8$\pm$13.7 & 16.66$\pm$0.18 & 16.90$\pm$0.35\\
221.8908818 & 07 Jan 2021 & 51 / 0.042 & 2005.9 & 222 & 2.79 & -61.9$\pm$13.1 & 16.53$\pm$0.18 & 16.95$\pm$0.35\\
222.9017998 & 08 Jan 2021 & 51 / 0.154 & 2005.9 & 229 & 2.61 & 5.3$\pm$12.6 & 16.64$\pm$0.18 & 17.42$\pm$0.35\\
\hline
886.0732816 & 02 Nov 2022 & 124 / 0.758 & 2005.9 & 293 & 1.94 & 62.8$\pm$10.5 & 16.69$\pm$0.18 & 17.35$\pm$0.35\\
890.0809514 & 06 Nov 2022 & 125 / 0.203 & 2005.9 & 260 & 2.27 & -87.1$\pm$13.1 & 17.04$\pm$0.18 & 17.23$\pm$0.35\\
891.0872272 & 07 Nov 2022 & 125 / 0.315 & 2005.9 & 269 & 2.21 & -145.6$\pm$14.1 & 16.47$\pm$0.18 & 17.37$\pm$0.35\\
892.0863444 & 08 Nov 2022 & 125 / 0.426 & 2005.9 & 251 & 2.36 & -102.2$\pm$13.7 & 16.37$\pm$0.18 & 17.69$\pm$0.35\\
893.0766032 & 09 Nov 2022 & 125 / 0.536 & 2005.9 & 259 & 2.19 & 8.6$\pm$12.8 & 17.24$\pm$0.18 & 17.72$\pm$0.35\\
895.0942424 & 11 Nov 2022 & 125 / 0.760 & 2005.9 & 234 & 2.48 & 55.1$\pm$13.4 & 16.49$\pm$0.18 & 16.78$\pm$0.35\\
896.0049589 & 12 Nov 2022 & 125 / 0.861 & 2005.9 & 288 & 1.96 & 6.7$\pm$11.4 & 17.07$\pm$0.18 & 17.52$\pm$0.35\\
899.0341013 & 15 Nov 2022 & 126 / 0.197 & 2005.9 & 257 & 2.27 & -52.4$\pm$12.7 & 16.98$\pm$0.18 & 17.61$\pm$0.35\\
900.1417289 & 16 Nov 2022 & 126 / 0.320 & 2005.9 & 247 & 2.35 & -129.7$\pm$13.6 & 16.12$\pm$0.18 & 16.68$\pm$0.35\\
901.0751357 & 17 Nov 2022 & 126 / 0.423 & 2005.9 & 253 & 2.40 & -69.4$\pm$12.4 & 16.69$\pm$0.18 & 17.13$\pm$0.35\\
902.1372620 & 18 Nov 2022 & 126 / 0.541 & 2005.9 & 200 & 5.50 & 40.7$\pm$31.4 & 16.97$\pm$0.18 & 17.48$\pm$0.35\\
903.1129681 & 19 Nov 2022 & 126 / 0.650 & 2005.9 & 273 & 2.21 & 62.3$\pm$11.4 & 16.65$\pm$0.18 & 17.14$\pm$0.35\\
905.0698972 & 21 Nov 2022 & 126 / 0.867 & 2005.9 & 213 & 2.72 & -3.2$\pm$14.2 & 16.93$\pm$0.18 & 17.15$\pm$0.35\\
906.0749340 & 22 Nov 2022 & 126 / 0.978 & 2005.9 & 231 & 2.50 & -35.6$\pm$13.1 & 16.56$\pm$0.18 & 16.86$\pm$0.35\\
915.0466839 & 01 Dec 2022 & 127 / 0.974 & 2005.9 & 287 & 2.03 & -38.6$\pm$14.6 & 17.42$\pm$0.18 & 17.52$\pm$0.35\\
918.0619499 & 04 Dec 2022 & 128 / 0.309 & 2005.9 & 265 & 2.23 & -115.2$\pm$18.2 & 16.97$\pm$0.18 & 18.62$\pm$0.35\\
945.9827166 & 01 Jan 2023 & 131 / 0.408 & 2005.9 & 252 & 2.31 & -156.4$\pm$12.3 & 16.98$\pm$0.18 & 18.14$\pm$0.35\\
946.9982611 & 02 Jan 2023 & 131 / 0.520 & 2005.9 & 257 & 2.28 & -39.1$\pm$11.7 & 16.71$\pm$0.18 & 17.06$\pm$0.35\\
947.9232632 & 03 Jan 2023 & 131 / 0.623 & 2005.9 & 250 & 2.24 & 82.7$\pm$10.3 & 16.69$\pm$0.18 & 17.13$\pm$0.35\\
948.9900594 & 04 Jan 2023 & 131 / 0.741 & 2005.9 & 257 & 2.26 & 107.0$\pm$11.3 & 16.84$\pm$0.18 & 17.02$\pm$0.35\\
949.9809414 & 05 Jan 2023 & 131 / 0.851 & 2005.9 & 246 & 2.39 & 39.0$\pm$12.2 & 16.86$\pm$0.18 & 16.96$\pm$0.35\\
952.9327461 & 08 Jan 2023 & 132 / 0.179 & 2005.9 & 226 & 2.61 & -87.7$\pm$13.8 & 16.88$\pm$0.18 & 16.79$\pm$0.35\\
953.9811468 & 09 Jan 2023 & 132 / 0.295 & 2005.9 & 242 & 2.50 & -168.1$\pm$12.8 & 16.39$\pm$0.18 & 16.79$\pm$0.35\\
955.9584385 & 11 Jan 2023 & 132 / 0.515 & 2005.9 & 250 & 2.33 & -32.4$\pm$11.7 & 17.20$\pm$0.18 & 17.41$\pm$0.35\\
956.8910560 & 12 Jan 2023 & 132 / 0.618 & 2005.9 & 260 & 2.23 & 70.2$\pm$11.7 & 16.76$\pm$0.18 & 17.06$\pm$0.35\\
957.9338770 & 13 Jan 2023 & 132 / 0.734 & 2005.9 & 267 & 2.16 & 117.3$\pm$11.8 & 16.88$\pm$0.18 & 16.93$\pm$0.35\\
\hline
\end{tabular}
\label{tab:lgs}
\end{table*}

\begin{table*} 
\small
\caption[]{Same as Table~\ref{tab:lgs} for our ESPaDOnS observations of CI~Tau in season 2020.  The peak SNR (per 2.3~\kms\ pixel) is obtained in the $I$ band. 
The last two columns list the \Bl\ values and RVs derived from the narrow components of the \caii\ IRT lines.  } 
\hspace{-6mm}
\begin{tabular}{cccccccccc} 
\hline
BJD        & date & c / $\phi$ & t$_{\rm exp}$ & SNR & $\sigma_V$            & \Bl\ LSD &  RV LSD   & \Bl\ IRT &  RV IRT \\ 
(2459000+) &      &            &   (s)        & ($I$) & ($10^{-4} I_c$)      & (G)      &  (\kms)   & (G)      &  (\kms) \\  
\hline
174.9434800 & 21 Nov 2020 & 45 / 0.832 & 4800.0 & 192 & 3.10 & -3.6$\pm$16.1 & 16.89$\pm$0.25 & -364.3$\pm$47.7 & 16.17$\pm$0.50 \\
178.8809500 & 25 Nov 2020 & 46 / 0.269 & 4800.0 & 180 & 3.43 & 88.0$\pm$19.7 & 16.52$\pm$0.25 & -253.1$\pm$77.2 & 16.26$\pm$0.50 \\
179.9065100 & 26 Nov 2020 & 46 / 0.383 & 4800.0 & 172 & 3.48 & 126.7$\pm$19.3 & 16.65$\pm$0.25 & -89.7$\pm$78.7 & 17.61$\pm$0.50 \\
180.8594600 & 27 Nov 2020 & 46 / 0.488 & 4800.0 & 206 & 2.76 & 119.7$\pm$16.1 & 16.59$\pm$0.25 & -549.4$\pm$127.5 & 17.16$\pm$0.50 \\
181.9250500 & 28 Nov 2020 & 46 / 0.607 & 4800.0 & 198 & 3.03 & 50.9$\pm$14.8 & 16.44$\pm$0.25 & -171.9$\pm$61.1 & 17.14$\pm$0.50 \\
183.0145900 & 29 Nov 2020 & 46 / 0.727 & 4800.0 & 205 & 2.90 & 15.5$\pm$13.0 & 16.53$\pm$0.25 & -464.8$\pm$74.6 & 17.24$\pm$0.50 \\
184.9605100 & 01 Dec 2020 & 46 / 0.943 & 4800.0 & 216 & 2.70 & 30.8$\pm$12.3 & 17.36$\pm$0.25 & -522.8$\pm$62.9 & 18.72$\pm$0.50 \\
185.8857900 & 02 Dec 2020 & 47 / 0.046 & 4800.0 & 186 & 3.22 & 23.1$\pm$15.8 & 16.49$\pm$0.25 & -414.2$\pm$92.8 & 18.60$\pm$0.50 \\
186.8910700 & 03 Dec 2020 & 47 / 0.158 & 4800.0 & 205 & 2.81 & 78.7$\pm$12.5 & 16.74$\pm$0.25 & -375.0$\pm$69.6 & 17.84$\pm$0.50 \\
187.8450500 & 04 Dec 2020 & 47 / 0.264 & 4800.0 & 221 & 2.63 & 42.3$\pm$14.5 & 16.78$\pm$0.25 & -225.5$\pm$57.0 & 17.57$\pm$0.50 \\
188.8595300 & 05 Dec 2020 & 47 / 0.376 & 4800.0 & 152 & 3.98 & 96.8$\pm$21.6 & 16.39$\pm$0.25 & -249.4$\pm$93.8 & 17.30$\pm$0.50 \\
189.9672600 & 06 Dec 2020 & 47 / 0.499 & 4800.0 & 208 & 2.84 & 85.6$\pm$13.4 & 16.89$\pm$0.25 & -204.5$\pm$59.7 & 17.02$\pm$0.50 \\
190.8734100 & 07 Dec 2020 & 47 / 0.600 & 4800.0 & 223 & 2.64 & 55.9$\pm$13.8 & 16.90$\pm$0.25 & -510.2$\pm$79.1 & 15.49$\pm$0.50 \\
\hline
\end{tabular}
\label{tab:lge}
\end{table*}

\section{ZDI fit to Stokes $I$ and $V$ ESPaDOnS and SPIRou profiles in 2020}
\label{sec:appB}

\begin{figure*}
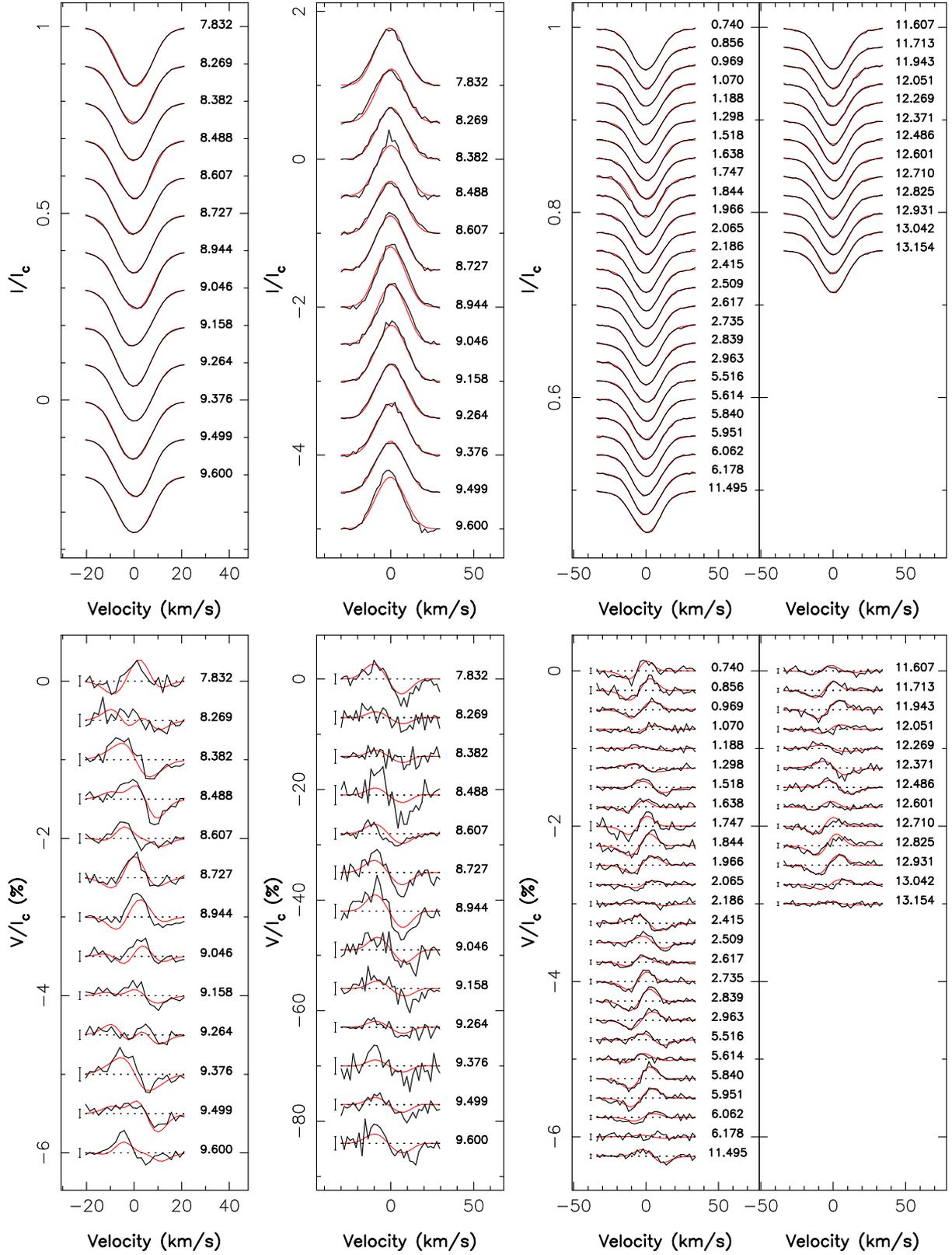

\centerline{\includegraphics[scale=0.6,angle=-90]{fig/citau2-fiti1e.ps}\hspace{2mm}\includegraphics[scale=0.6,angle=-90]{fig/citau2-fiti2e.ps}\hspace{2mm}\includegraphics[scale=0.6,angle=-90]{fig/citau2-fitis.ps}\vspace{2mm}} 
\centerline{\includegraphics[scale=0.6,angle=-90]{fig/citau2-fitv1e.ps}\hspace{2mm}\includegraphics[scale=0.6,angle=-90]{fig/citau2-fitv2e.ps}\hspace{2mm}\includegraphics[scale=0.6,angle=-90]{fig/citau2-fitvs.ps}} 
\caption[]{Observed (thick black line) and modelled (thin red line) LSD Stokes $I$ (top panels) and $V$ (bottom) profiles of CI~Tau, for the photospheric lines in the ESPaDOnS domain (left), the \caii\ IRT lines (middle) and 
the photospheric lines in the SPIRou domain (right) for our 2020 observing season.  Rotation cycles (counting from 38, see Table~\ref{tab:lge}) are indicated to the right of all LSD profiles, while $\pm$1$\sigma$ error bars 
are added to the left of Stokes $V$ signatures. } 
\label{fig:fite}
\end{figure*}

We show in Fig.~\ref{fig:fite} the Stokes $I$ and $V$ LSD profiles for the three sets of lines monitored in season 2020, along with the ZDI fit in the case $\delta=0.4$ (see Sec.~\ref{sec:zd2}).

\section{2D periodograms of the 1083~nm \hei\ and \pab\ lines} 
\label{sec:appC}

\begin{figure*}
\hbox{\includegraphics[scale=0.33,bb=0 0 760 1170]{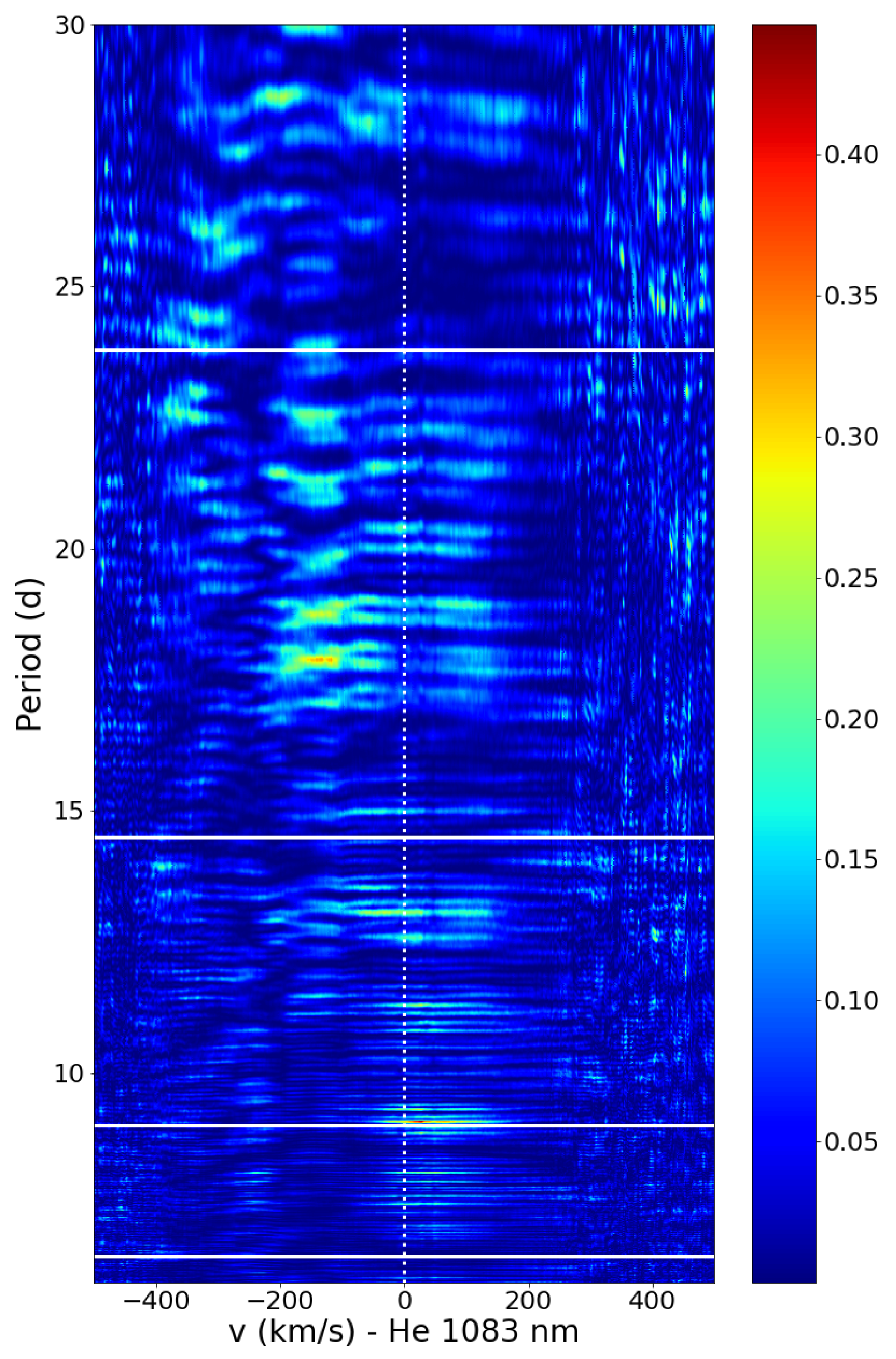}\hspace{2mm}\includegraphics[scale=0.33,bb=0 0 760 1170]{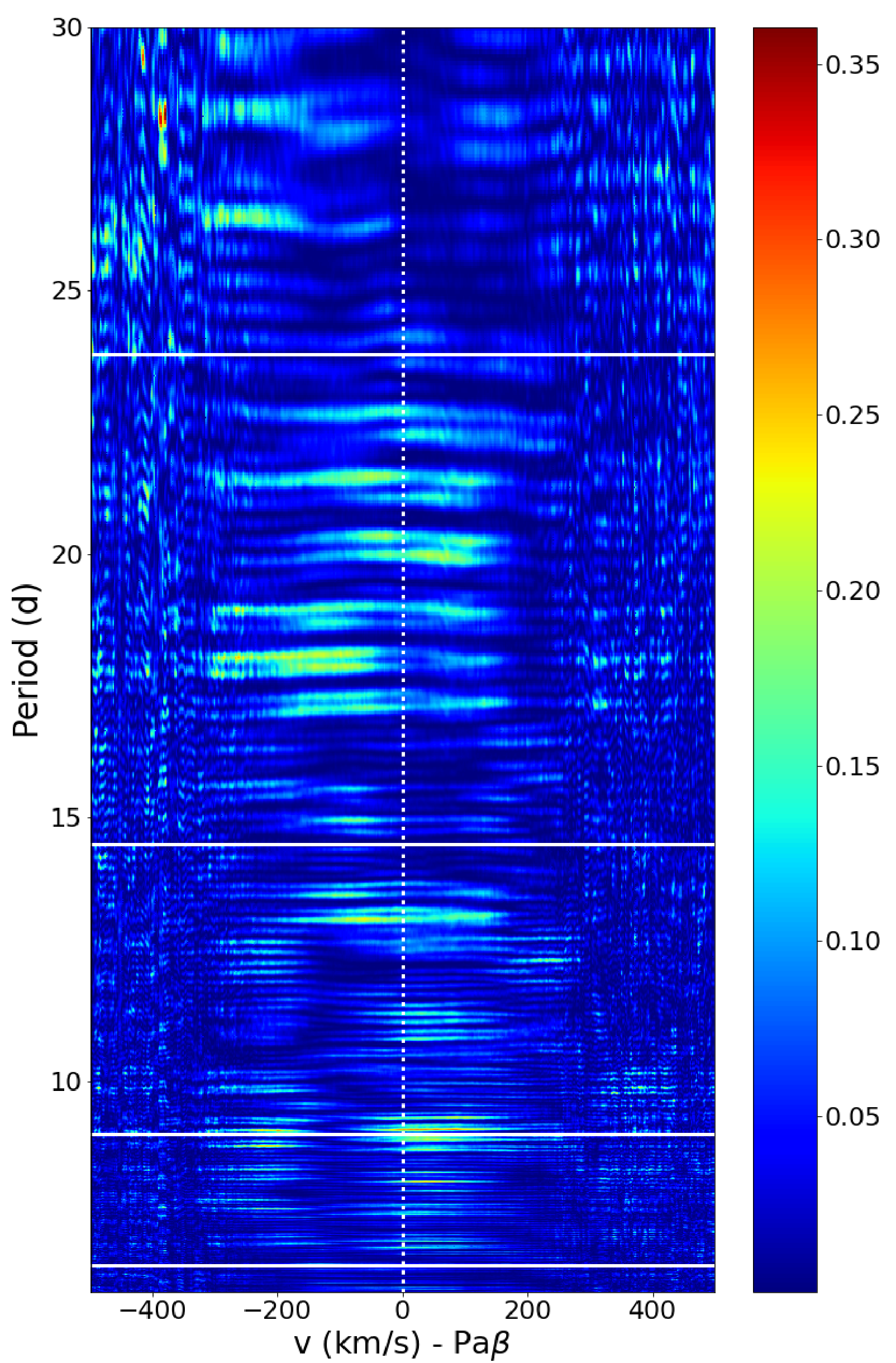}} 
\caption[]{2D periodograms of the 1083~nm \hei\ (left) and \pab\ (right) lines for periods ranging from 6 to 30~d.  The white horizontal lines depict periods the rotation period derived from our \Bl\ data (9.01~d),
the period of the RV signal detected from CO bandhead lines (23.86~d), and the additional periods inferred from the K2 light-curve \citep[i.e., 6.6 and 14.3~d][]{Biddle18,Biddle21}}.  
\label{fig:perf}
\end{figure*}

We show in Fig.~\ref{fig:perf} the 2D periodograms of the 1083~nm \hei\ (left) and \pab\ (right) lines for periods ranging from 6 to 30~d. Only weak signals are observed apart from that at the rotation period.  

\section{Profiles of the \hal\ and \hbe\ lines} 
\label{sec:appD}

\begin{figure*}
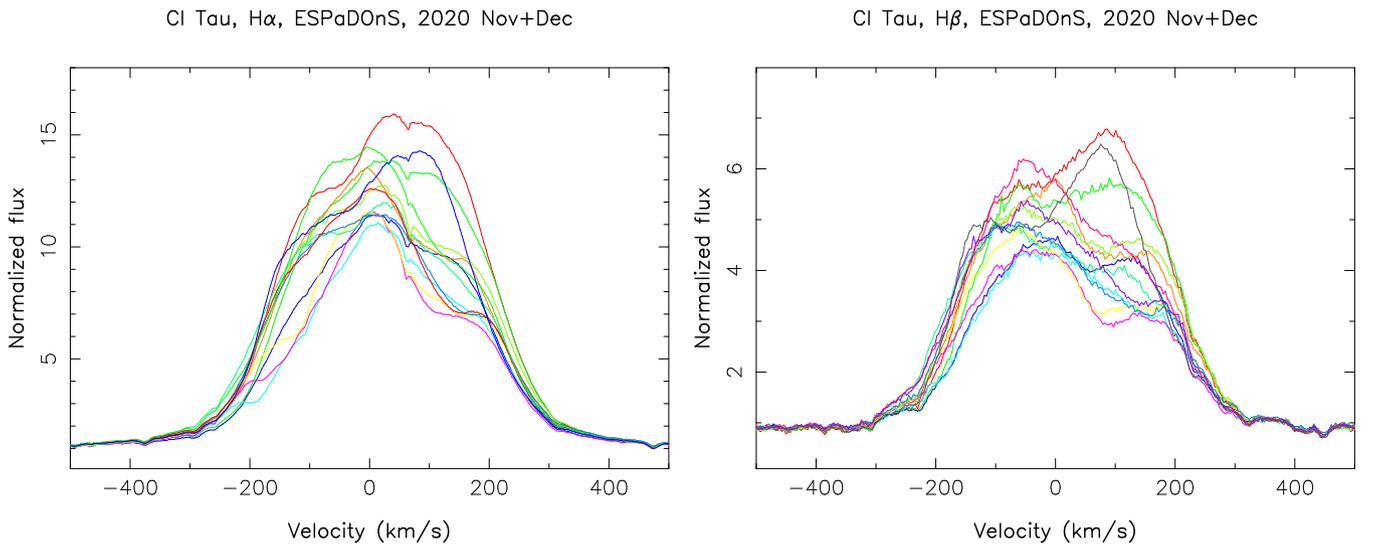

\hbox{\includegraphics[scale=0.38,angle=-90]{fig/citau2-hal.ps}\hspace{3mm}\includegraphics[scale=0.38,angle=-90]{fig/citau2-hbe.ps}} 
\caption[]{Stacked profiles of the  \hal\ (left) and \hbe\ (right) line profiles of CI~Tau during our ESPaDOnS run of late 2020.}  
\label{fig:bal}
\end{figure*}

We show in Fig.~\ref{fig:bal} the 13 profiles (stacked on top of each other) of the \hal\ (left) and \hbe\ (right) lines in CI~Tau, as recorded by ESPaDOnS in 2020 November and December (see Table~\ref{tab:lge}).  

\bsp    
\label{lastpage}
\end{document}